\documentclass[10pt,twocolumn]{article}

\ifdefined\pdfgentounicode
  \input{glyphtounicode}
\fi
\usepackage{cmap}
\usepackage[T1]{fontenc}
\usepackage[utf8]{inputenc}
\usepackage[a4paper, top=2.0cm, bottom=2.4cm, left=1.5cm, right=1.5cm, columnsep=0.65cm]{geometry}
\usepackage{amsmath}
\usepackage{amssymb}
\usepackage{amsthm}
\usepackage{booktabs}
\usepackage{hyperref}
\usepackage{microtype}
\usepackage{graphicx}
\graphicspath{{figures/}{bioeth-beacon-bundle/paper/figures/}}
\usepackage{url}
\usepackage{xcolor}
\usepackage{array}
\usepackage{tabularx}
\usepackage{titlesec}
\usepackage{caption}
\usepackage{enumitem}
\usepackage{natbib}
\usepackage{algorithm}
\usepackage{algpseudocode}
\usepackage{multirow}
\usepackage{fancyhdr}

\hypersetup{
  colorlinks=true,
  linkcolor=blue!60!black,
  citecolor=blue!60!black,
  urlcolor=blue!60!black,
  pdftitle={bioETH-Beacon: A Confidential On-Chain Genomic Beacon with Encrypted Counts, Filters, and Bounded Noise over a Fully Homomorphic EVM},
  pdfauthor={Christos Galanopoulos; Kimon Antonios Provatas; Ilias Georgakopoulos-Soares},
  pdfsubject={A research prototype for confidential GA4GH Beacon aggregate-count queries over a fully homomorphic EVM},
  pdfkeywords={confidential genomic querying; GA4GH Beacon; fully homomorphic encryption; fhEVM; smart contracts; membership-inference mitigation; anti-probing noise; privacy-preserving genomics}
}
\titleformat{\section}{\normalfont\large\bfseries\scshape}{\thesection}{1em}{}
\titleformat{\subsection}{\normalfont\normalsize\bfseries}{\thesubsection}{1em}{}
\titleformat{\subsubsection}{\normalfont\normalsize\itshape\bfseries}{\thesubsubsection}{1em}{}

\setlist[itemize]{leftmargin=*, topsep=2pt, itemsep=1pt}
\setlist[enumerate]{leftmargin=*, topsep=2pt, itemsep=1pt}

\newcolumntype{Y}{>{\raggedright\arraybackslash}X}

\newtheorem{invariant}{Invariant}

\title{\textbf{bioETH-Beacon: A Confidential On-Chain Genomic Beacon\\[2pt]
       with Encrypted Counts, Filters, and Bounded Noise over\\[2pt]
       a Fully Homomorphic EVM}}

\author{%
  \begin{minipage}{0.94\textwidth}
  \centering
  {\large\textbf{Christos Galanopoulos$^{1}$, Kimon Antonios Provatas$^{1,*}$}}\\[1pt]
  {\large\textbf{Ilias Georgakopoulos-Soares$^{1,*}$}}\\[4pt]
  \normalsize $^{1}$Division of Pharmacology and Toxicology, College of Pharmacy, The University of Texas at Austin, Dell\\
  \normalsize Pediatric Research Institute, Austin, TX, USA\\[4pt]
  \small $^{*}$Corresponding authors: \texttt{kap4722@my.utexas.edu}; \texttt{ilias@austin.utexas.edu}\\[4pt]
  \begin{tabular}{c}
  \small\textit{Confidential Genomic Querying $\cdot$ GA4GH Beacon $\cdot$ FHE $\cdot$ Smart Contracts}\\
  \small\textit{Anti-Probing Noise $\cdot$ Membership-Inference Mitigation}
  \end{tabular}
  \end{minipage}
}
\date{}

\begin{document}
\sloppy
\maketitle
\thispagestyle{fancy}

\begin{figure*}[t]
  \centering
  \includegraphics[width=0.96\linewidth]{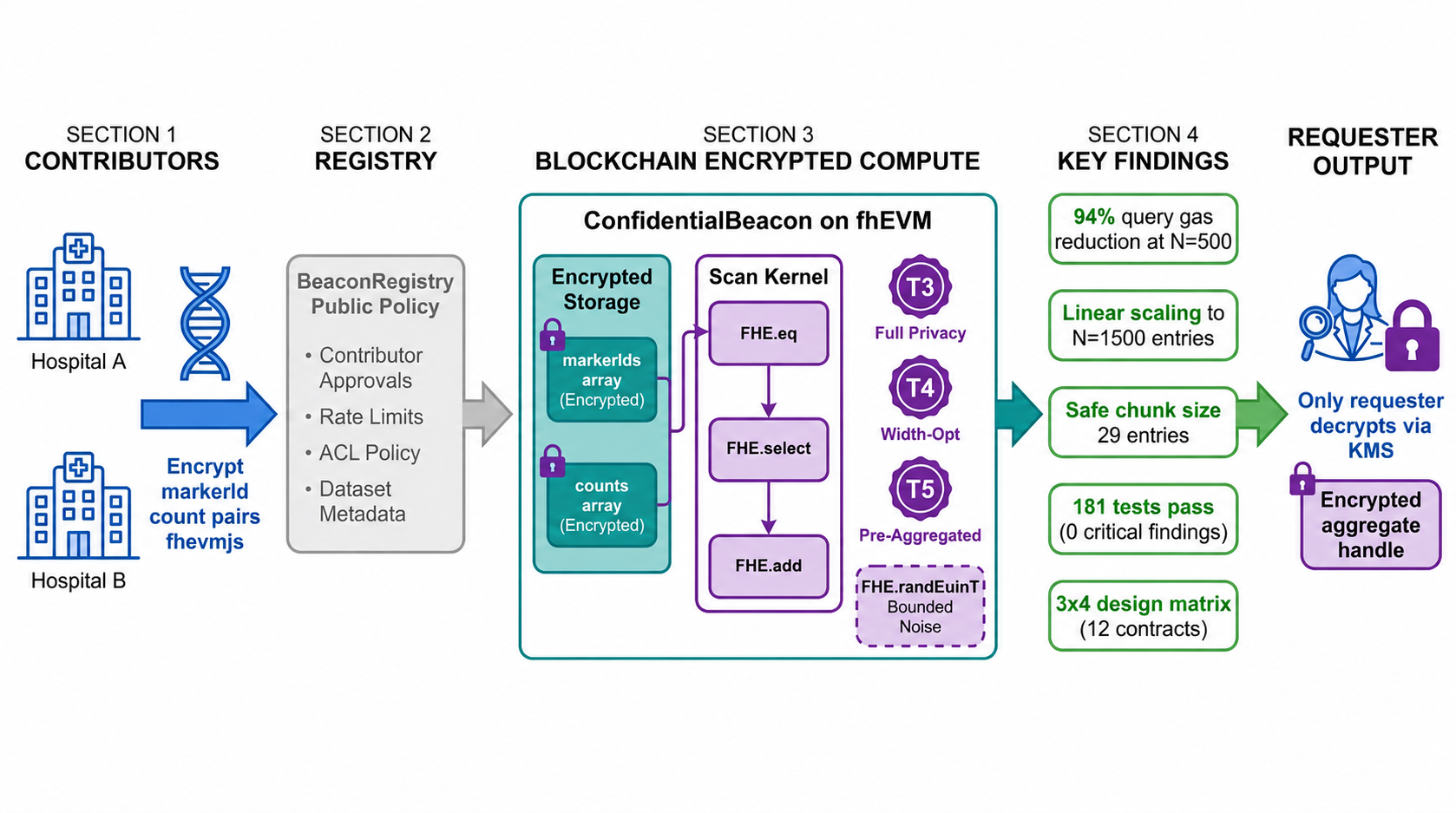}
  \caption{\textbf{Graphical Abstract.}  bioETH-Beacon executes a Beacon-style
  aggregate-count primitive entirely in the encrypted domain on a fully homomorphic EVM.
  Hospitals upload encrypted marker-count entries; an authorized researcher submits a
  single encrypted marker query; the contract returns an encrypted aggregate without exposing
  raw genomic data, contributor counts, or the queried marker.  Optional bounded
  noise on genotype and G1 query paths increases anti-probing cost but is not a
  differential-privacy guarantee.  Across a $3\!\times\!4$ tier-by-query-family grid,
  measured local Hardhat EVM query gas in the fixed-$M{=}20$ tier-comparison
  benchmark at $N{=}500$ entries ranges from $24.9$M (reference encrypted
  scan) to $1.45$M (pre-aggregated, $-94\%$).  All gas figures in this paper
  are local EVM-gas only; on a live fhEVM coprocessor the HCU adder is
  comparable to or larger than the EVM floor.}
  \label{fig:graphical_abstract}
\end{figure*}

\begin{abstract}
\noindent
The Global Alliance for Genomics and Health (GA4GH) Beacon protocol lets
researchers ask whether a genomic variant has been observed in a participating
cohort and, when supported, receive aggregate variant-level counts.  As Beacon
networks grow, two privacy risks remain: host institutions can see plaintext
queries, and repeated rare-variant queries can support membership-inference
attacks.  We present bioETH-Beacon, a smart-contract prototype that runs the
Beacon ``aggregate count'' query over encrypted data on a fully homomorphic Ethereum Virtual Machine
(fhEVM).  Hospitals upload encrypted marker-count entries, authorized
researchers submit encrypted marker queries, and the contract returns an
encrypted answer that is released, via the off-chain key-management service,
only to the requester named in the contract's on-chain ACL.  The design is organized
as a $3\!\times\!4$ tier-by-query-family grid spanning genotype, sex, age, and phenotype
queries, with tiers that trade stronger confidentiality for lower query cost.
It also includes a G1 multi-filter contract for one encrypted four-way query.
For genotype and G1 paths, the prototype can add bounded on-chain noise to make
probing attacks harder; single-filter sex, age, and phenotype queries currently
return exact encrypted counts.  This noise is an anti-probing feature, not a
formal differential-privacy guarantee.  Experiments on synthetic panels derived
from a Polygenic Score (PGS) catalog show the expected scaling behavior,
clarify practical chunk-size limits, and demonstrate that pre-aggregation can
substantially reduce query gas when public marker presence is an acceptable
trade-off.  The paper also states tier-qualified security invariants for the
implemented contract suite.  Overall, bioETH-Beacon is a research prototype for
confidential Beacon-style genomic querying, not a production-ready clinical
deployment.
\end{abstract}

\section*{Key Points}
\begin{itemize}
  \item bioETH-Beacon implements the GA4GH Beacon ``aggregate count'' query directly on
        fhEVM: marker identifiers, counts, and result handles remain encrypted,
        and only authorized requesters can retrieve the plaintext answer
        through the off-chain key-management service.
  \item A $3\!\times\!4$ tier-by-query-family grid, three execution tiers crossed with four query
        families, presents the design space in a readable form, from fully
        encrypted evaluation to faster pre-aggregated lookup and encrypted
        multi-filter queries.
  \item Bounded on-chain query-noise is implemented on $4$ of the $13$
        deployed contract paths (genotype T3/T4/T5 and the G1 multi-filter);
        the $9$ single-axis sex, age, and phenotype filter contracts currently
        return exact encrypted counts and require governed requester access.
        On the supported paths the noise sample is drawn inside the FHE
        coprocessor, so coordinator-chosen zero-noise injection is prevented
        by construction; natural zero-noise draws still occur with probability
        $1/B$.
  \item Benchmarks clarify the practical tradeoff: smaller chunks stay within
        fhEVM depth limits, while pre-aggregation substantially lowers query
        gas by revealing only public marker presence.
  \item Tier-qualified security invariants are checked across the prototype's
        unit, integration, scaling, filter, noise-injection, and G1 test suites.
\end{itemize}

\section{Introduction}
\label{sec:intro}

The Global Alliance for Genomics and Health (GA4GH) Beacon protocol
\citep{fiume2019federated} is a deliberately simple primitive for federated
genomic discovery: it determines whether a particular sequence variant has been
observed in a participating institution's cohort, with optional allele-frequency
metadata in Beacon v1 and richer variant-level quantitative summaries such as
matching variant count, call count, and sample count in later Beacon models
\citep{rambla2022beacon}.  Beacon networks span biobanks, hospital networks,
and research consortia, and the protocol has expanded from genomic-variant
discovery to a richer phenotype-clinical model that combines variant presence
and count-style
responses with filters over patient attributes such as reported sex, age band,
or phenotype ontology terms.

\textbf{The double-disclosure problem.}
Despite its popularity, deployed Beacons leave two privacy gaps largely
unaddressed.  First, the host institution observes every query in plaintext:
which researcher submitted it, which variant it concerns, and how many carriers
were returned.  Second, the researcher receives the count in plaintext, and
repeated queries against rare variants enable a membership-inference attack
\citep{shringarpure2015privacy} that can confirm the presence of a target
individual in the cohort.  Both gaps can be partially mitigated organizationally
 by trusting the host, by access governance, by rate limiting, but
neither is closed by the protocol itself.

\textbf{The opportunity.}
Fully homomorphic encryption (FHE), and in particular the bounded-integer TFHE
scheme \citep{chillotti2020tfhe} as deployed by the Zama fhEVM coprocessor
\citep{zamafhevm}, allows arithmetic over encrypted operands without
decryption.  Combined with a programmable blockchain, this opens a path where
the Beacon ``count'' primitive runs inside an immutable smart contract that
neither the host nor the requester needs to trust as a designated \emph{compute}
evaluator: the host and chain see public requester, dataset, event-timing, and
metadata information plus encrypted query contents.  Result decryption itself is
still delegated to an off-chain key-management service (KMS) that holds the
threshold-shared decryption key (Section~\ref{sec:security} discusses this
trust assumption explicitly), and only the requester named in the contract's
\texttt{FHE.allow} call can retrieve the plaintext aggregate through that
service.

\textbf{Our contribution.}
We present bioETH-Beacon, an open-source confidential Beacon protocol
that implements the GA4GH aggregate-count query directly on fhEVM.  The
contributions are:

\begin{enumerate}
  \item A two-layer contract architecture comprising a public \texttt{BeaconRegistry}
        that holds policy, contributor approvals, and rate limits, and an
        encrypted \texttt{ConfidentialBeacon} family that holds all genomic state and
        performs the encrypted scan.  No plaintext genomic counts appear on-chain, and T3/T4 also hide marker and
        filter values; T5 intentionally exposes marker presence as a
        documented trade-off (Section~\ref{sec:design}).
  \item A $3\!\times\!4$ tier-by-query-family grid that crosses three execution tiers (\texttt{T3}
        reference, \texttt{T4} width-optimized, \texttt{T5} pre-aggregated) with four query
        families (genotype, reported sex group, age band, phenotype term), plus
        a multi-filter contract evaluating a four-way conjunctive predicate in a
        single encrypted query (Section~\ref{sec:matrix}).
  \item An encrypted scan kernel that runs an oblivious
        \texttt{FHE.eq}\,+\,\texttt{FHE.select}\,+\,\texttt{FHE.add} chain over every entry
        without early exit, with chunk-size advisory pinned by an empirically
        measured \texttt{HCUTransactionDepthLimitExceeded} ceiling
        (Section~\ref{sec:kernel}).
  \item An on-chain bounded uniform noise injection that adds width-matched
        \texttt{FHE.randEuintT} noise inside the FHE coprocessor before result
        release ($T{=}64$ for T3/G1 and $T{=}32$ for T4/T5).  Unlike off-chain
        noise schemes, the realized noise value is hidden from the coordinator
        until the block is mined, preventing coordinator-chosen zero-noise
        injection (Section~\ref{sec:noise}).
  \item An empirical evaluation on synthetic panels derived from the PGS Catalog
        \citep{lambert2021pgscatalog}, spanning 27 to 1,500 uploaded entries and panel sizes from 16 to
        100,000 markers, with measured gas, transaction-count, wall-time,
        and HCU-per-query-entry profiles across the evaluated tiers (Section~\ref{sec:eval}).
  \item A security analysis built around eleven explicit invariants
        (Section~\ref{sec:security}) across the implemented prototype contract
        suite.
\end{enumerate}

The system is implemented as a Solidity contract suite with Hardhat tests,
Sepolia-only pending T4 ceiling probes, and benchmark profiles.  The measured
numbers reported here are sourced from the artifacts generated by the
accompanying benchmark pipeline.

\section{Methods}
\label{sec:methods}

\subsection{GA4GH Beacon Protocol}

The GA4GH initiative \citep{ga4gh2016genomics} launched Beacon as part of a
broader federated ecosystem.  The Beacon protocol
\citep{fiume2019federated,rambla2022beacon} exposes a
small set of queries over a hidden cohort: a v1 Beacon is centered on genomic
variant requests with yes/no responses and optional allele-frequency metadata,
whereas later Beacon models expose richer quantitative fields such as matching
variant count, call count, and sample count \citep{rambla2022beacon}.  Beacon v2 generalizes
the model with structured results and richer filtering support
(\emph{filteringTerms}) over patient attributes such as ontology terms or
demographic categories.  The protocol is intentionally narrow in semantics,
there is no record-level disclosure, only aggregates, but the threat model
is delegated entirely to the host's operational practice.
Two attacks documented in the literature shape the rest of this paper.
Shringarpure and Bustamante \citep{shringarpure2015privacy} demonstrated that an
adversary querying a Beacon over rare variants from a target genome can use a
likelihood-ratio test to infer membership in the cohort after repeated
queries.  McLaren \emph{et al.}\ \citep{mclaren2016privacy} demonstrated a
clinical model using homomorphic encryption and secure two-party protocols.  In
this framework, the storage and processing unit (SPU) receives encrypted marker
requests, while the client can include dummy variants so that request size does
not directly reveal the nature of the test.

\subsection{TFHE on a Programmable Blockchain}

The fhEVM \citep{zamafhevm} extends the Ethereum Virtual Machine with a TFHE
\citep{chillotti2020tfhe} coprocessor that operates over encrypted unsigned
integers up to $256$ bits.  Solidity authors manipulate opaque handles
(\texttt{euint32}, \texttt{euint64}, etc.) via library calls such as
\texttt{FHE.eq}, \texttt{FHE.select}, \texttt{FHE.add}, and width-specific random
samplers such as \texttt{FHE.randEuint32} and \texttt{FHE.randEuint64}.  The
coprocessor charges a \emph{Homomorphic Computation Unit} (HCU) cost adder on
top of EVM gas.  In the documented testnet configuration used for this
evaluation, two binding limits apply per transaction: a global HCU budget of
$20$M units and a sequential-depth budget of $5$M units, which constrains the
longest chain of dependent FHE operations \emph{(e.g., a chain of 29 dependent
\texttt{FHE.add(euint64)} operations consumes $29 \times 133{,}000 =
3{,}857{,}000$ of the 5,000,000-unit depth budget)} \citep{zamafhevmdocs}.
Access control on each encrypted handle is mediated by an on-chain ACL: a
handle is \emph{persisted} for a contract via \texttt{FHE.allowThis}, granted to a
specific address via \texttt{FHE.allow(handle, addr)}, or made world-readable via
\texttt{FHE.makePubliclyDecryptable} \emph{(e.g., per-entry counts stay at
\texttt{allowThis} forever; only the final aggregate transitions to
\texttt{allow(handle, requester)}; \texttt{makePubliclyDecryptable} is never
invoked anywhere in the protocol)}.  Decryption itself is delegated to an
off-chain key-management service (KMS) using an EIP-712 (Ethereum Improvement Proposal 712) signed user request.

Three properties of this stack shape the bioETH-Beacon design.  First,
fhEVM integer arithmetic is exact over unsigned integers, with no floating-point
approximation --- well suited to the count semantics of Beacon queries.
Second, the sequential-depth limit binds before the global HCU limit for any
operation that accumulates into a single ciphertext, dictating chunk-size
choices for our encrypted scan.  Third, the ACL is the only barrier between an
encrypted handle and its plaintext: the discipline of never calling
\texttt{makePubliclyDecryptable} on an aggregate count is the single most important
security property in the entire protocol.

\subsection{Membership Inference and Output Minimization}

Two complementary mitigations against the Shringarpure--Bustamante attack are
used in the supported parts of the design.  The first is \emph{output minimization}: the
aggregate count is never released as a plaintext value or to any party other
than the originating requester; the chain itself stores only opaque ciphertext
handles.  The second is \emph{bounded noise} on genotype and G1 query paths:
the released count is perturbed by an integer drawn uniformly from
$\{0,\ldots,B-1\}$, where $B$ is a power-of-two committed on-chain at dataset
creation.  Section~\ref{sec:noise} gives the mechanism, expected upward bias
$(B-1)/2$, and calibration guidance.

\section{System Design}
\label{sec:design}

\begin{figure}[t]
  \centering
  \includegraphics[width=\linewidth]{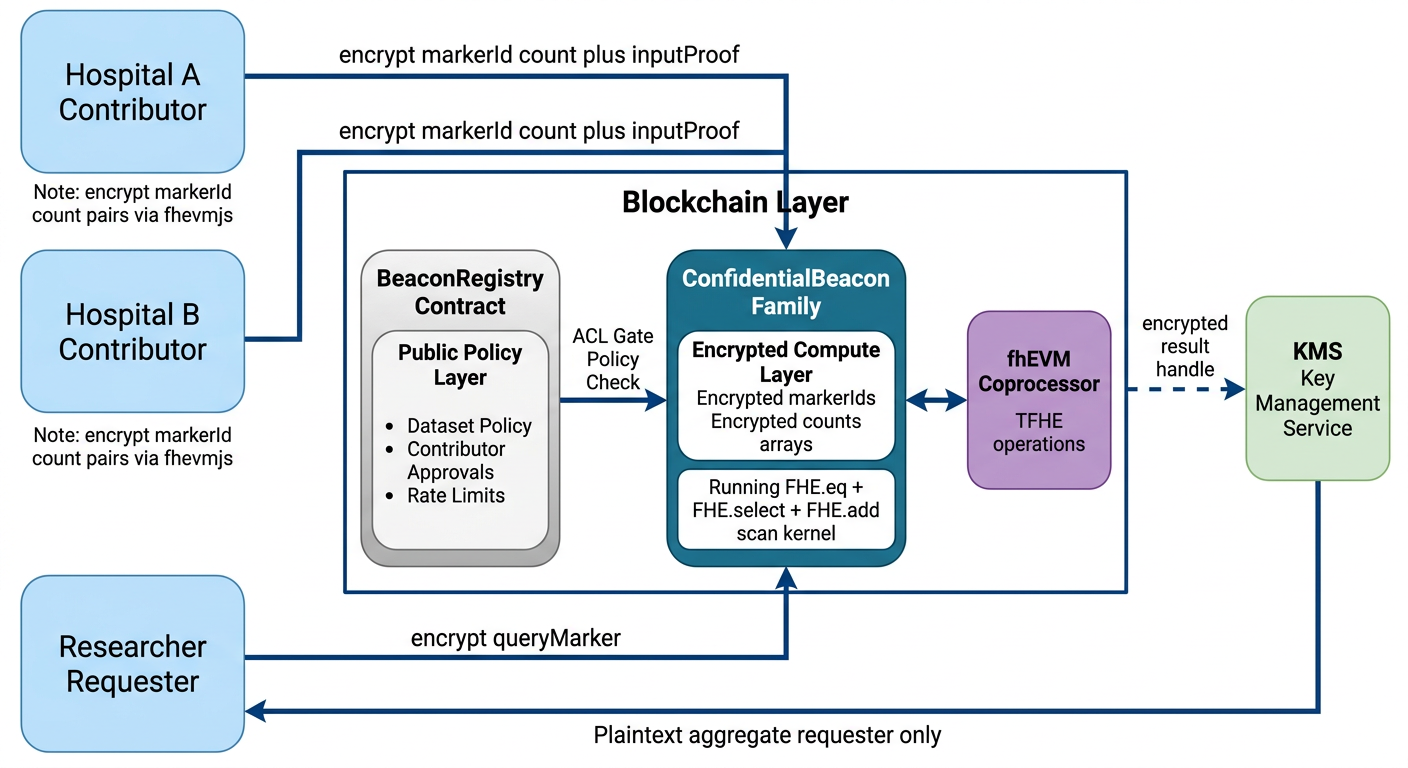}
  \caption{\textbf{System architecture.} Approved hospitals encrypt
  marker--count pairs off-chain using \texttt{fhevmjs} and submit them with an input
  proof.  The public \texttt{BeaconRegistry} gates uploads and queries; the encrypted
  \texttt{ConfidentialBeacon} family stores and scans encrypted state and emits a
  requester-private result handle decrypted via the off-chain KMS.}
  \label{fig:architecture}
\end{figure}

\subsection{Architecture}

bioETH-Beacon splits cleanly into a public policy layer and an encrypted
compute layer.  Figure~\ref{fig:architecture} sketches the data flow.  The
public layer (\texttt{BeaconRegistry}) holds dataset metadata, the manifest URI
and integrity commitment of the shared marker dictionary, contributor approval
state, the per-(dataset, requester) rate-limit policy, and a retirement flag,
and contains no encrypted state.  The encrypted layer
(\texttt{ConfidentialBeacon}\{\texttt{T3},\,\texttt{T4},\,\texttt{T5}\} and the
filter and multi-filter variants) holds the encrypted marker--count arrays
per dataset, runs the encrypted scan kernel, and on \texttt{finalizeQuery}
calls \texttt{FHE.allow(handle, requester)} so the requester can recover the
plaintext count via an EIP-712 user-decrypt request to the off-chain KMS.
Both layers are deployed as ordinary Solidity contracts on fhEVM; the
registry is consulted from the compute layer through pure view calls, which
reduces the cross-contract reentrancy surface.

\subsection{Marker Canonicalization}

Variant strings are normalized to opaque $32$-bit identifiers off-chain.  Each
contributor and querier agree on a shared dictionary anchored by the manifest
URI and integrity commitment in \texttt{BeaconRegistry}.  The on-chain identifier of
a variant $v$ is
\begin{equation}
  \begin{split}
  \mathrm{markerId}(v) ={}&
  \mathrm{sha256}\!\big(\mathrm{genomeBuild}\,\Vert\,\mathrm{dictVersion}\,\Vert{} \\
  &\mathrm{norm}\,\Vert\,v\big)_{0:32},
  \end{split}
  \label{eq:markerid}
\end{equation}
where the subscript $_{0:32}$ denotes the leading $32$ bits of the SHA-$256$
digest over the canonical concatenation.  The first four bytes are interpreted as an unsigned big-endian integer.
The consortium manifest workflow should reject dictionaries that produce duplicate \texttt{markerId} values before deployment.
This narrows the encrypted comparison from \texttt{euint64} to \texttt{euint32} and
halves the per-entry \texttt{FHE.eq} HCU; in exchange, the birthday-bound
collision probability becomes non-trivial above $\sim\!10^{4}$ distinct
markers ($\approx 1.2\%$ at $10{,}000$) and reaches $\approx 39\%$ at $2^{16}$.
The operating regime is therefore deterministic reject-and-remap below
$\sim\!10^{4}$ markers, reject-with-explicit-collision-table up to
$\sim\!6.5\!\times\!10^{4}$ markers (where remapping becomes impractical), and
migration to the wider \texttt{euint128} encoding above that.
Section~\ref{sec:limitations} discusses the population-scale mitigation path.  The dictionary itself is
never read on-chain; integrity is anchored off-chain through the manifest
URI and digest registered at \texttt{createDatasetShell} time.

\subsection{State Machines}

\begin{figure}[t]
  \centering
  \includegraphics[width=\linewidth]{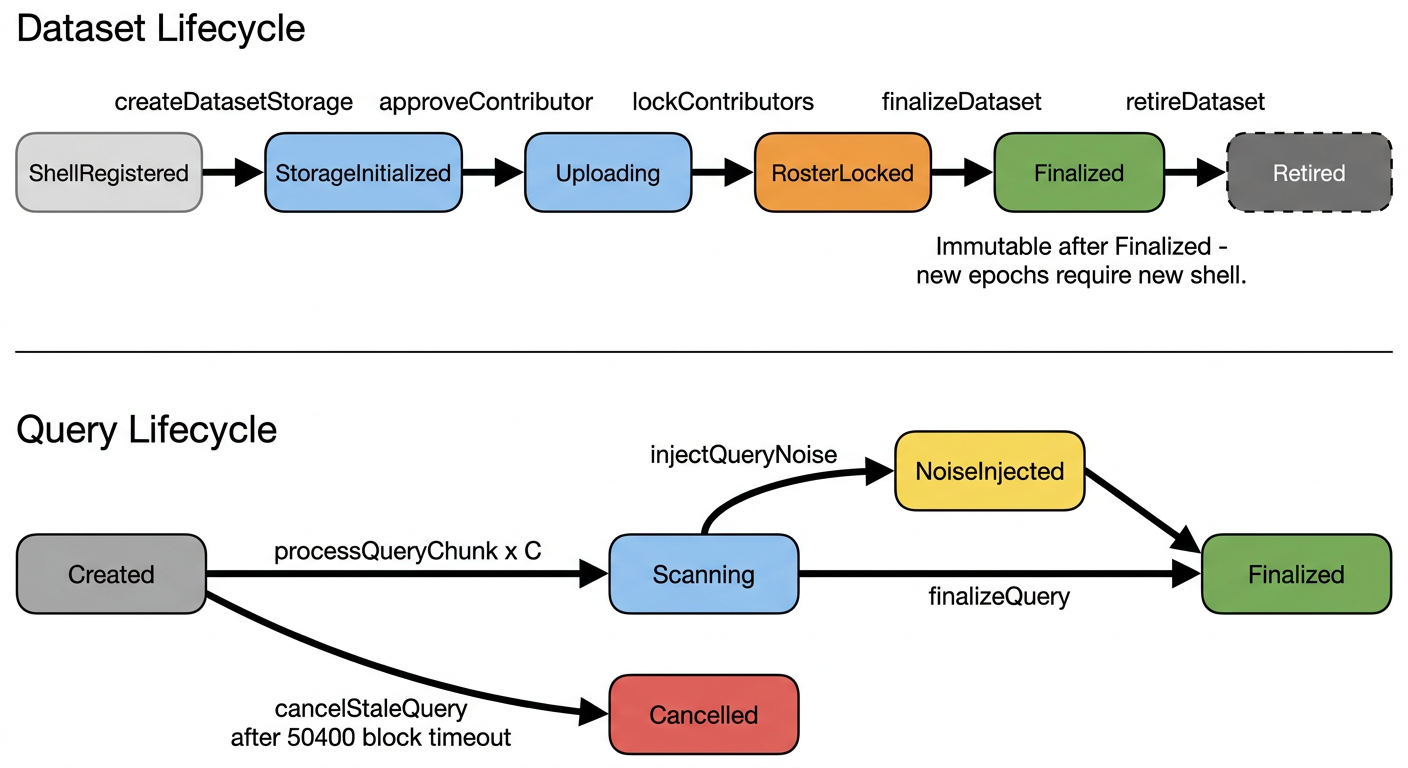}
  \caption{\textbf{Dataset and query lifecycles.}  Datasets become immutable at
  finalization and can only be replaced by registering a new shell.  Queries are
  monotone: \texttt{processQueryChunk} only advances \texttt{nextEntryIndex}, so
  permissionless relayers cannot stall a query.}
  \label{fig:lifecycle}
\end{figure}

Two state machines run independently.  The \emph{dataset lifecycle}
(Figure~\ref{fig:lifecycle}, top) advances from \texttt{ShellRegistered}
\emph{(e.g., dataset \texttt{0x4a...} is registered with \texttt{entryCount =
1500} and dictionary commitment \texttt{0x2c...} before any encrypted data is
uploaded)} via
\texttt{StorageInitialized} \emph{(e.g., the owner calls
\texttt{createDatasetStorage} so subsequent uploads have allocated encrypted
arrays to write into)} and
\texttt{Uploading} to \texttt{RosterLocked} \emph{(e.g., after
\texttt{lockContributors}, no further institutions may join this dataset epoch)} and finally
\texttt{Finalized}; a coordinator may transition a finalized dataset to
\texttt{Retired}, after which no new queries are accepted.  Datasets are immutable
once finalized; new epochs require a fresh shell registration with a new
\texttt{datasetId} \emph{(e.g., the 2026 Q1 and 2026 Q2 cohorts are separate
epochs, each bound to its own shell and roster)}.  The \emph{query lifecycle} (Figure~\ref{fig:lifecycle},
bottom) advances from \texttt{Created} through one or more \texttt{processQueryChunk}
transactions to a fully scanned state, then through optional
\texttt{injectQueryNoise} before \texttt{finalizeQuery} grants the aggregate handle
to the requester.  Stale queries can be cancelled permissionlessly after a
$50{,}400$-block ($\approx 7$-day) timeout, ensuring that abandoned queries
cannot grief storage indefinitely \emph{(e.g., a query abandoned after two scan
chunks can be reclaimed by any address calling \texttt{cancelStaleQuery} once
50,400 blocks have elapsed)}.

\subsection{Roles, Authorization, and Rate Limits}

Three roles are recognized: a \emph{coordinator} owns a dataset and controls
contributor approval, rate-limit policy, and retirement \emph{(e.g., the
consortium steering committee that approves member hospitals and sets the
per-requester rate-limit policy)}; a \emph{contributor} is an approved
institution that uploads encrypted marker--count chunks \emph{(e.g., North
Regional Hospital, after being added to the roster, uploads eight encrypted
marker--count pairs across two chunks)}; and a \emph{requester} submits queries
and is the only party able to decrypt results \emph{(e.g., a researcher at
address \texttt{0x9f...} who submits an encrypted marker query and later decrypts
the returned aggregate handle through the KMS)}.  A \emph{relayer} role is implicit \emph{(e.g., a
consortium-funded service that watches \texttt{QueryCreated} events and pays gas
to drive \texttt{processQueryChunk} calls to completion on the requester's
behalf)}: \texttt{processQueryChunk} is
permissionless because it can only advance a query monotonically, so any
gas-paying party can help a query complete without affecting correctness.
Rate limits are enforced \texttt{(datasetId, requester)}-wise in
\texttt{BeaconRegistry} using \texttt{block.number} windows rather than wall-clock
time, making the limiter robust to the $\pm 15$-second timestamp drift that
block proposers can introduce \emph{(e.g., a 10-query budget over a
7,200-block window --- $\approx 24$ hours at 12-second blocks --- caps a single
requester to at most ten successful \texttt{createQuery} calls before the window
slides)}.  Open-access exact-count querying is intended
only for low-sensitivity datasets: address-based quotas are operational
throttles, not identity-level privacy controls, and sensitive genomic cohorts
should require governed requester approval rather than open exact-count access.

\section{\texorpdfstring{The $3\!\times\!4$ tier-by-query-family grid}{The 3x4 tier-by-query-family grid}}
\label{sec:matrix}

The bioETH-Beacon contract suite is organized as a $3\!\times\!4$
tier-by-query-family grid.  Three execution tiers parameterise the cost/privacy trade-off of the
encrypted scan, and four query families select the filter axes recognized by
the Beacon v2 schema \emph{(e.g., the sex family exposes five plaintext-cardinality
buckets --- unknown, female, male, other, withheld --- and contributors upload an
encrypted bucket id alongside each marker)}.  A separate
\emph{multi-filter} contract (G1, ``first-generation multi-axis''), described
below, collapses the four families into a single encrypted four-way
conjunctive query and is the thirteenth deployed contract.  The T3/T4/T5
numbering is historical: T1 and T2 were internal optimization milestones, and
T3/T4/T5 are the maintained contract families.

\emph{Size symbols used throughout.}  $P$ = panel/dictionary marker count
($P{=}16$, $100$, $1{,}000$, etc.); $N$ = number of uploaded sparse non-zero
entries scanned by T3/T4 (a function of the per-hospital density and the
panel size); $M$ = number of T5 unique-marker slots scanned per query (a
fixed-at-deployment parameter, $M \ll N$ in the regime where T5 is
worthwhile).  The P$16$ worked example has $P{=}16$ and $N{=}31$; $M$ is not
defined for T3/T4.

\begin{figure}[t]
  \centering
  \includegraphics[width=\linewidth]{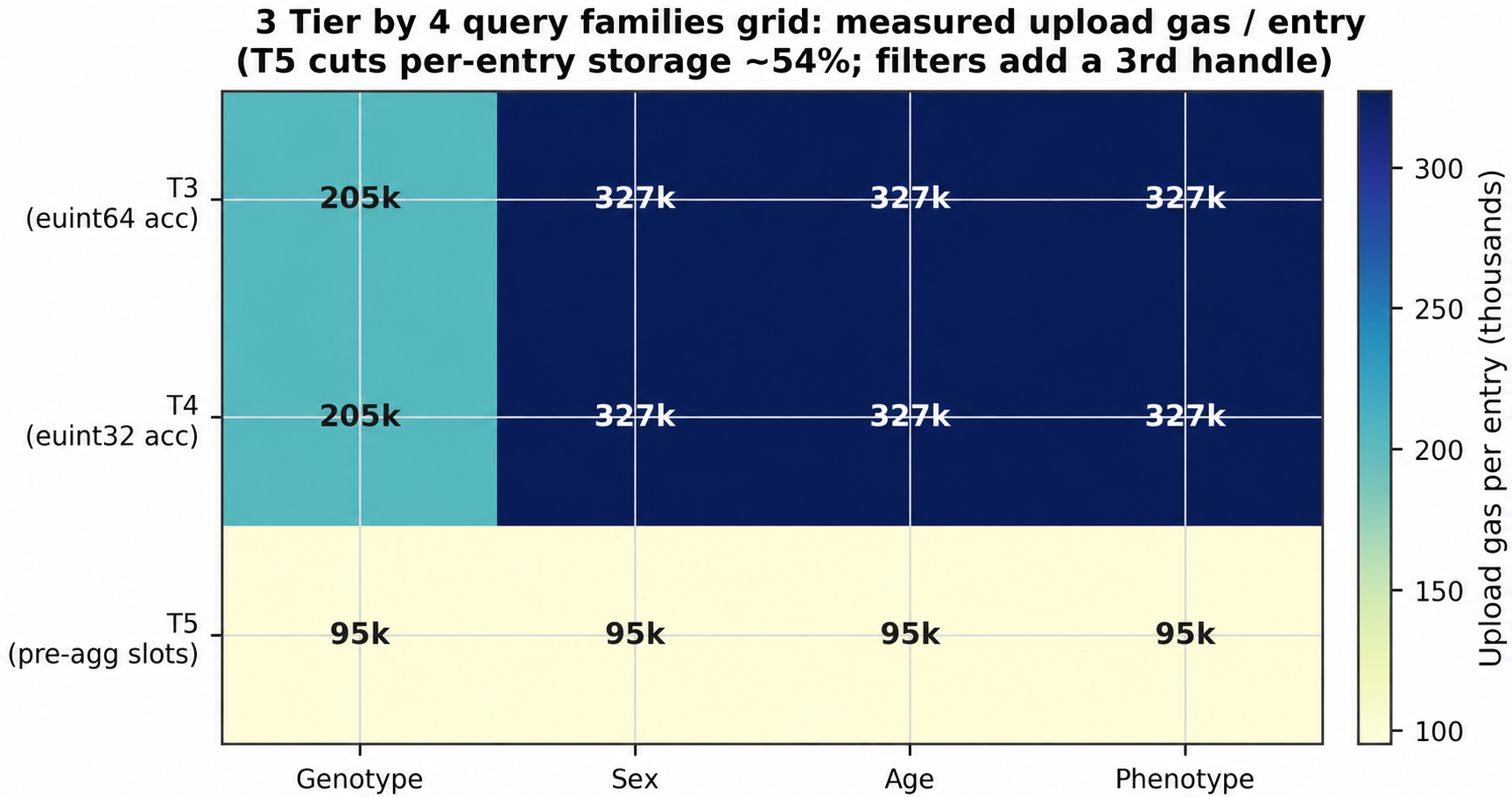}
  \caption{\textbf{The $3\!\times\!4$ tier-by-query-family grid.}  Twelve contracts cover the
  cross-product of tiers and query families; a thirteenth (G1) collapses the
  four families into a single encrypted four-way conjunctive query.}
  \label{fig:design_matrix}
\end{figure}

Table~\ref{tab:tier-semantics} summarizes the tier
semantics; Figure~\ref{fig:design_matrix} overlays the measured cost deltas at
$N{=}500$ and $M{=}20$ from Section~\ref{sec:eval}.

\subsection{Execution Tiers}

\textbf{T3 --- reference encrypted scan.}  The base tier stores encrypted
$32$-bit marker IDs and encrypted $64$-bit counts and evaluates the scan kernel
$\texttt{eq} \to \texttt{select} \to \texttt{add}$ on every entry.  Marker IDs, filter values,
and counts are all encrypted.  Query cost is $\Theta(N)$ in the number of
uploaded entries.

\textbf{T4 --- ciphertext-width optimization.}  T4 narrows the count and
accumulator from \texttt{euint64} to \texttt{euint32}.  Privacy is unchanged
(everything is still encrypted), but deployments must ensure that every true
aggregate plus the maximum optional noise $(B{-}1)$ fits in \texttt{uint32}.
Under that precondition, \texttt{FHE.add} drops from $133$k to $95$k
sequential-depth HCU per step, lifting the chunk-size ceiling from $37$ to
$52$ entries on a live coprocessor.  Local Hardhat receipts do not include
the HCU adder, so T4's gain is visible only on a live network; we report the
HCU model analytically in Section~\ref{sec:eval}.

\textbf{T5 --- pre-aggregated slot mode.}  T5 changes the storage layout
entirely: the coordinator declares a fixed set of $M$ marker slots at
\texttt{createDatasetStorage}, and contributors upload \texttt{(slotIndex, encCount)}
pairs that accumulate into the slot's encrypted counter \emph{(e.g., a T5
dataset with $M = 20$ slots exposes \texttt{slotMarkerIds[0..19]} in plaintext;
contributors upload \texttt{(slotIndex, encCount)} pairs that the contract adds
homomorphically into \texttt{slotCounts[slotIndex]})}.  Queries scan
$M$ slots rather than $N$ rows, yielding $\Theta(M)$ query cost regardless of
how many entries have been uploaded.  The trade-off is explicit: slot indices
are plaintext, revealing which markers each contributor covers (though counts
remain encrypted).  The query marker remains encrypted in the current T5 scan
implementation, but the public slot list and contributor slot indices reveal
dataset and contributor marker presence.  T5 is intended for dense, query-heavy
workloads where that marker-presence pattern is not sensitive.

\begin{table}[t]
  \centering
  \caption{\textbf{Execution-tier semantics.}  Privacy axes describe what is
  hidden from an on-chain observer.  Cost class is asymptotic query cost.}
  \label{tab:tier-semantics}
  \small
  \begingroup
  \setlength{\tabcolsep}{3pt}
  \begin{tabularx}{\columnwidth}{@{}p{0.09\columnwidth}p{0.13\columnwidth}p{0.12\columnwidth}p{0.14\columnwidth}Y@{}}
    \toprule
    Tier & Marker priv. & Count priv. & Cost class & When to use \\
    \midrule
    T3 & Yes & Yes & $\Theta(N)$ & Full privacy; use for any dataset \\
    T4 & Yes & Yes & $\Theta(N)$ & Full privacy; lower cost on live networks; safe unless per-marker count exceeds $\sim$4 billion \\
    T5 & No & Yes & $\Theta(M)$ & Fastest; marker IDs are public; best for dense panels with many queries \\
    \bottomrule
  \end{tabularx}
  \endgroup
\end{table}

\subsection{Query Families}

The four families add a categorical filter dimension to the encrypted scan
kernel.  In each case the per-entry kernel becomes
\begin{equation*}
\begin{aligned}
  m_i &= \texttt{FHE.eq}(\mathit{markerIds}[i],\ q.\mathit{queryMarker}),\\
  f_i &= \texttt{FHE.eq}(\mathit{filterIds}[i],\ q.\mathit{queryFilter}),\\
  \texttt{isMatch}_i &= \texttt{FHE.and}(m_i, f_i),
\end{aligned}
\end{equation*}
adding two operations and lifting the global per-entry HCU from $277$k to
$\sim\!342$k for T3 filter variants.  The sequential-depth ceiling remains
dominated by the accumulator chain, but global HCU headroom falls by
$\sim\!23\%$ relative to the base T3 path.

\textbf{Genotype} (no extra filter).  The base family; \texttt{T3}, \texttt{T4}, \texttt{T5}
genotype contracts answer ``how many carriers of marker $m$?''

\textbf{Reported sex group} (5 buckets: unknown, female, male, other,
withheld).  The lowest-cardinality filter, used to demonstrate the filter
extension pattern with minimal overhead.

\textbf{Age band} (7 buckets: unknown, 0--17, 18--29, 30--39, 40--49,
50--59, 60+).  Identical kernel structure to sex; bucket boundaries match
common epidemiological strata.

\textbf{Phenotype} (configurable, 1--$1000$ Human Phenotype Ontology (HPO)
or NCI Thesaurus (NCIT) terms).  The
highest-cardinality filter, intended for ontology-anchored disease queries.
Cardinality is fixed at contract construction so deployments can match the
relevant ontology slice without recompilation.

\textbf{Multi-filter conjunctive query (G1, ``first-generation multi-axis'').}
A separate contract (\texttt{ConfidentialBeaconMultiFilterT3}) accepts four
encrypted predicates --- marker, sex, age, phenotype --- in a single query and
returns a single encrypted aggregate that satisfies all four simultaneously \emph{(e.g., a single G1 query
for marker $\wedge$ sex = female $\wedge$ age = 40--49 $\wedge$ phenotype =
HP:0002664 returns one encrypted aggregate without ever computing intermediate
single-axis counts)}.  The per-entry kernel adds
$4\!\times\!\texttt{FHE.eq} + 3\!\times\!\texttt{FHE.and}$ for a total of $\sim\!472$k
HCU per entry, $+70\%$ over the base T3 cost.  The contract is strictly stronger, from a privacy standpoint, than issuing
four sequential single-dimension queries: no
intermediate count is ever computed or disclosed.

\subsection{Path-Property Reference}
\label{sec:path-table}

Table~\ref{tab:path-properties} consolidates the privacy axes, cost class,
and bounded-noise support of all $13$ deployed contracts.  It is the
canonical reference for which paths carry anti-probing noise (the
$4$ shaded rows) and which paths currently return exact encrypted counts
(the $9$ remaining rows).

\begin{table}[t]
  \centering
  \caption{\textbf{Deployed contract paths and properties.}
  ``Marker priv.'' and ``Count priv.'' indicate whether the marker identifier
  and the per-entry count remain encrypted on-chain (T5 exposes marker slot
  indices as a documented trade-off).  ``Noise'' indicates whether bounded
  on-chain query-noise (Section~\ref{sec:noise}) is implemented on that
  path.}
  \label{tab:path-properties}
  \scriptsize
  \begingroup
  \setlength{\tabcolsep}{3pt}
  \begin{tabularx}{\columnwidth}{@{}lYYccc@{}}
    \toprule
    Contract & Tier & Family & Marker & Count & Noise \\
              &      &        & priv.  & priv. &       \\
    \midrule
    \texttt{T3}              & T3 & genotype  & Yes & Yes & \textbf{Yes} \\
    \texttt{T4}              & T4 & genotype  & Yes & Yes & \textbf{Yes} \\
    \texttt{T5}              & T5 & genotype  & No  & Yes & \textbf{Yes} \\
    \texttt{SexT3}           & T3 & sex       & Yes & Yes & No \\
    \texttt{SexT4}           & T4 & sex       & Yes & Yes & No \\
    \texttt{SexT5}           & T5 & sex       & No  & Yes & No \\
    \texttt{AgeT3}           & T3 & age       & Yes & Yes & No \\
    \texttt{AgeT4}           & T4 & age       & Yes & Yes & No \\
    \texttt{AgeT5}           & T5 & age       & No  & Yes & No \\
    \texttt{PhenotypeT3}     & T3 & phenotype & Yes & Yes & No \\
    \texttt{PhenotypeT4}     & T4 & phenotype & Yes & Yes & No \\
    \texttt{PhenotypeT5}     & T5 & phenotype & No  & Yes & No \\
    \texttt{MultiFilterT3}   & T3 & G1 (4-way) & Yes & Yes & \textbf{Yes} \\
    \bottomrule
  \end{tabularx}
  \endgroup
  \par\smallskip
  {\scriptsize All contracts share the \texttt{ConfidentialBeacon} prefix
  (e.g., \texttt{ConfidentialBeaconSexT3}).  Cost class: $\Theta(N)$ for all
  T3 and T4 rows, $\Theta(M)$ for all T5 rows, $\Theta(N)$ for the G1
  multi-filter row.}
\end{table}

\section{Encrypted Scan Kernel}
\label{sec:kernel}

\subsection{Per-Entry Kernel}

\begin{figure}[t]
  \centering
  \includegraphics[width=\linewidth]{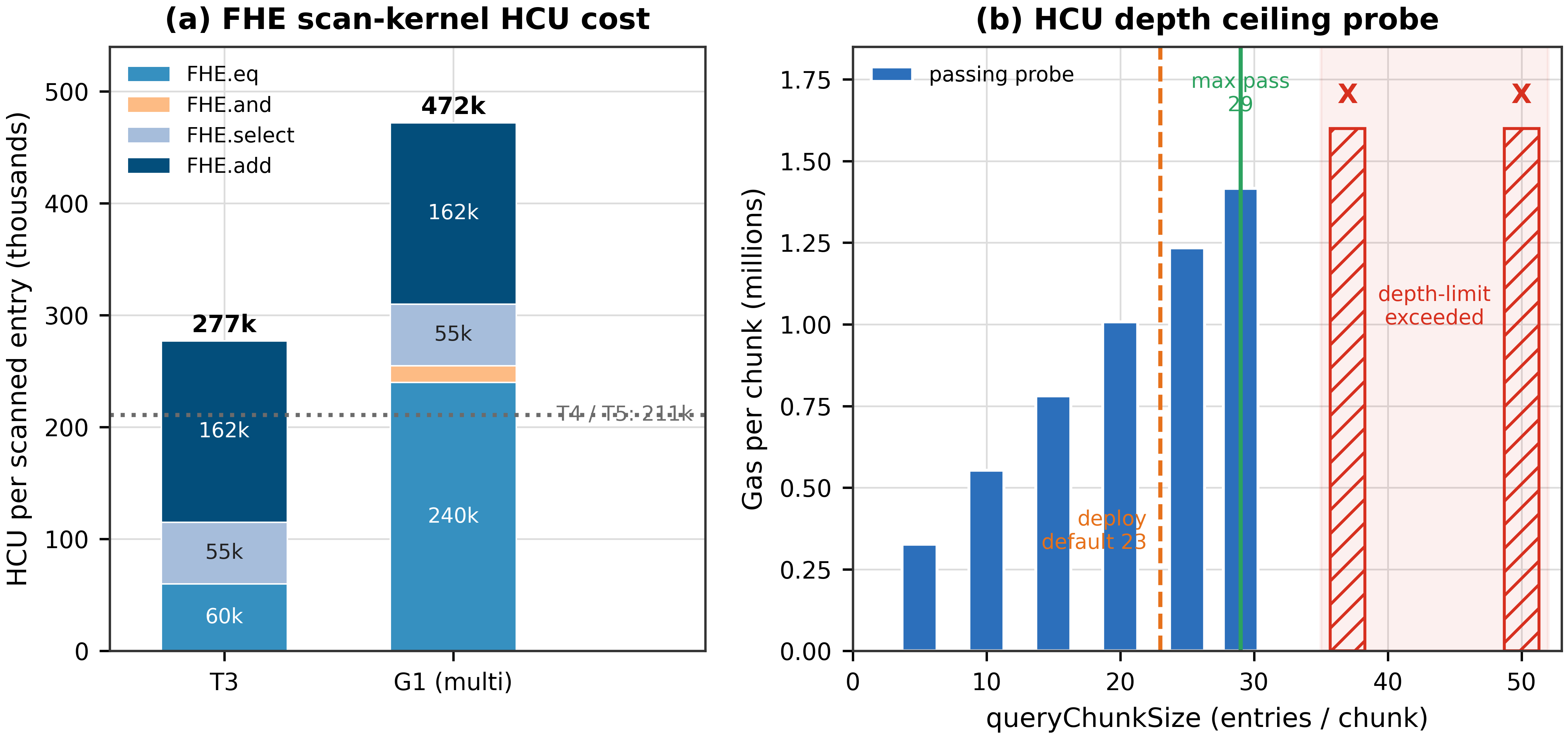}
  \caption{\textbf{Per-entry FHE scan flow.}  HCU figures refer to the
  $\texttt{euint32}\,\texttt{eq}$ + $\texttt{euint64}\,\texttt{add}$ reference path
  (T3).  The accumulator \texttt{add} dominates sequential depth and sets the
  chunk-size ceiling.}
  \label{fig:scan_kernel}
\end{figure}

The reference T3 scan kernel is a four-line oblivious computation:
\begin{equation*}
\begin{aligned}
  &\texttt{isMatch}   = \texttt{FHE.eq}(\texttt{markerIds}[i],\ \texttt{queryMarker}), \\
  &\texttt{zero}      = \texttt{FHE.asEuintT}(0), \\
  &\texttt{addend}    = \texttt{FHE.select}(\texttt{isMatch},\ \texttt{counts}[i],\ \texttt{zero}), \\
  &\texttt{acc}       = \texttt{FHE.add}(\texttt{acc},\ \texttt{addend}).
\end{aligned}
\end{equation*}
Both branches of the \texttt{FHE.select} are evaluated regardless of the encrypted
condition.  The kernel is explicitly free of early-exit branches: this is
security Invariant~\ref{inv:fullscan} in Section~\ref{sec:security}.
Figure~\ref{fig:scan_kernel} renders the per-entry data flow and shows where
the $133$k-HCU \texttt{FHE.add} step binds the sequential-depth budget.

\subsection{Chunking and the Depth Ceiling}

A single transaction cannot process arbitrarily many entries because
\texttt{FHE.add} forms a sequential chain into the accumulator.  At
$133{,}000$ sequential-depth HCU per step and a $5{,}000{,}000$ per-transaction
budget, the theoretical ceiling is
$\lfloor 5\!\times\!10^{6} / 133{,}000 \rfloor = 37$ entries per chunk.
We tested this empirically with Hardhat; chunk sizes
$5, 10, 20, 29$ all pass, while $37$ already fails with
\texttt{HCUTransactionDepthLimitExceeded}.  The repository
therefore uses three distinct chunk-size figures: $29$ is the
\emph{largest tested passing value} for \texttt{euint64} accumulators and is the
configuration used by the profile, benchmark, and worked-example runs in
Section~\ref{sec:eval}; $23$ is the \emph{deployment-manifest recommendation}
($\approx\!80\%$ of $29$, a $20\%$ safety margin); and $37$ is the
\emph{analytic add-chain ceiling} from
$\lfloor 5\!\times\!10^{6}/133{,}000\rfloor$, which the probe shows fails in
practice.  Tests in the range $30$--$36$ have not been run, so the true
empirical ceiling lies somewhere in that interval; the $23$/$29$/$37$ triad is
used consistently throughout this paper and is enforced by
Invariant~\ref{inv:chunksize} and the pre-deployment advisor
(\texttt{scripts/chunk-size-advisor.ts}).

For T4's narrower \texttt{euint32} accumulator the depth cost per step drops
to $95$k HCU, lifting the analytic ceiling to $\lfloor 5{,}000{,}000/95{,}000
\rfloor = 52$ and the recommended $80\%$ chunk to $41$.  The local Hardhat environment
enforces the same depth across types, so the T4 advantage is not visible in
local profiling --- only on a live network.

\subsection{Algorithms}

The chunked encrypted scan (T3/T4) and the pre-aggregated slot evaluation (T5)
are stated in Algorithms~\ref{alg:t3scan} and~\ref{alg:t5scan}.

\begin{algorithm}[t]
\caption{Chunked encrypted scan (T3/T4)}
\label{alg:t3scan}
\small
\begin{algorithmic}[1]
\Procedure{processQueryChunk}{$queryId$}
  \State $q \gets \mathit{queries}[queryId]$
  \State $start \gets q.\mathit{nextEntryIndex}$
  \State $end \gets \min(start + \texttt{queryChunkSize},\ N)$
  \State $acc \gets q.\mathit{aggregatedCount}$
  \For{$i \gets start$ \textbf{to} $end{-}1$}
    \State $m \gets \texttt{FHE.eq}(\mathit{markerIds}[i],\ q.\mathit{queryMarker})$
    \State $\delta \gets \texttt{FHE.select}(m,\ \mathit{counts}[i],\ \texttt{FHE.asEuintT}(0))$
    \State $acc \gets \texttt{FHE.add}(acc,\ \delta)$
  \EndFor
  \State $\texttt{FHE.allowThis}(acc)$
  \State $q.\mathit{aggregatedCount} \gets acc;\ q.\mathit{nextEntryIndex} \gets end$
\EndProcedure
\end{algorithmic}
\end{algorithm}

\begin{algorithm}[t]
\caption{Pre-aggregated slot evaluation (T5)}
\label{alg:t5scan}
\small
\begin{algorithmic}[1]
\Procedure{processQueryChunk}{$queryId$}
  \Comment{$M$ slots, $M \ll N$ assumed}
  \State $q \gets \mathit{queries}[queryId]$
  \State $start \gets q.\mathit{nextSlotIndex}$
  \State $end \gets \min(start + \texttt{queryChunkSize},\ M)$
  \State $acc \gets q.\mathit{aggregatedCount}$
  \State $zero \gets \texttt{FHE.asEuintT}(0)$
  \For{$j \gets start$ \textbf{to} $end{-}1$}
    \State $m \gets \texttt{FHE.eq}(\mathit{slotMarkerIds}[j],\ q.\mathit{queryMarker})$
    \State $\delta \gets \texttt{FHE.select}(m,\ \mathit{slotCounts}[j],\ zero)$
    \State $acc \gets \texttt{FHE.add}(acc,\ \delta)$
  \EndFor
  \State $\texttt{FHE.allowThis}(acc)$
  \State $q.\mathit{aggregatedCount} \gets acc;\ q.\mathit{nextSlotIndex} \gets end$
\EndProcedure
\end{algorithmic}
\end{algorithm}

\section{Bounded Query-Noise Injection}
\label{sec:noise}

\subsection{Mechanism}

\begin{figure}[t]
  \centering
  \includegraphics[width=\linewidth]{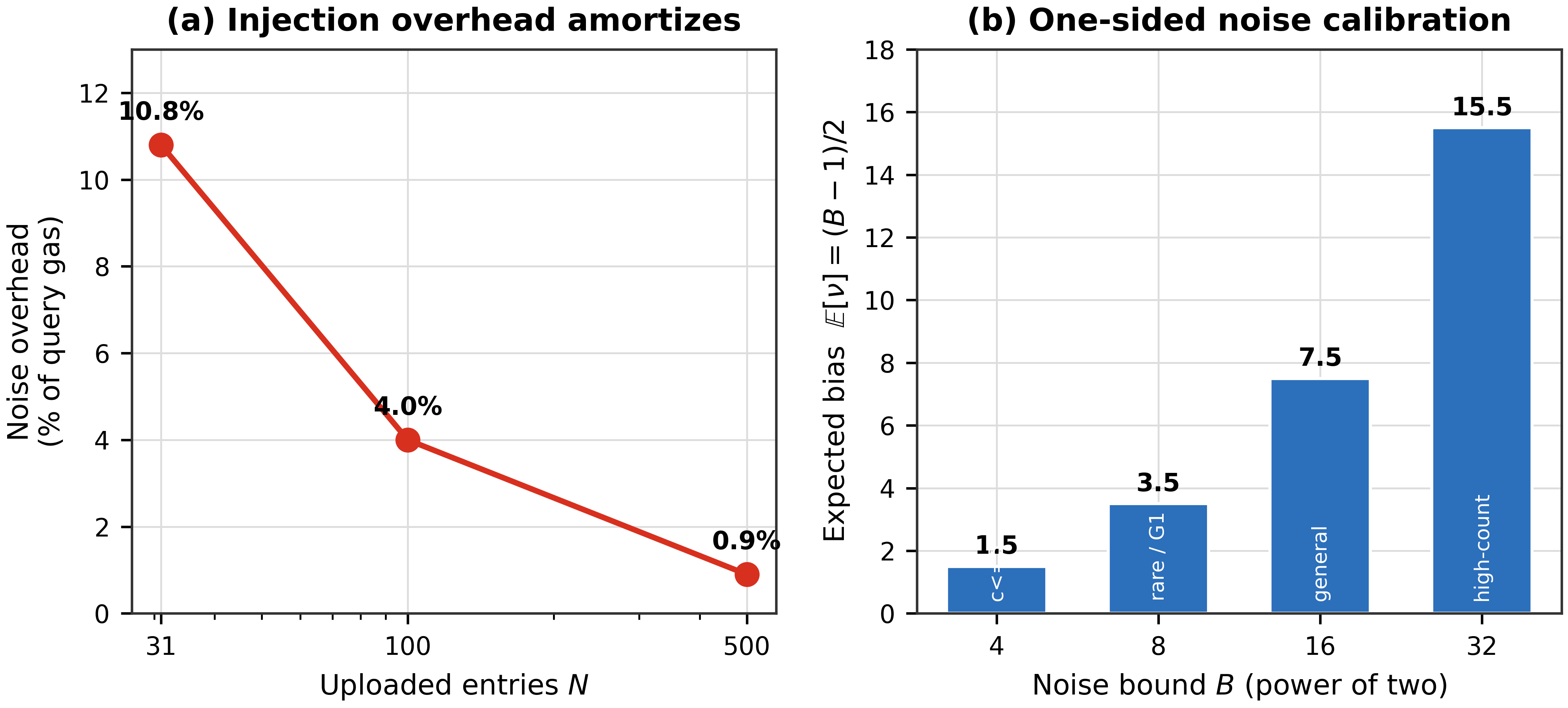}
  \caption{\textbf{Bounded on-chain noise injection.}  \emph{Left:} Overhead gas
  decreases as a fraction of total query gas as $N$ grows.  \emph{Right:} The
  noise sample $\nu$ is generated inside the FHE coprocessor and never exists in
  plaintext; the coordinator triggers the injection but cannot observe or
  substitute the sampled value before the block mines.}
  \label{fig:noise}
\end{figure}

The base protocol returns the exact encrypted aggregate.  Although the
ciphertext is requester-private, an adversarial requester can issue many
queries against rare variants and reconstruct cohort membership through a
likelihood-ratio test \citep{shringarpure2015privacy}.  Rate limits delay
but do not close this channel.  bioETH-Beacon adds an optional on-chain
bounded uniform anti-probing noise injection on supported query paths (genotype
T3/T4/T5 and G1 in the current implementation):
\begin{equation}
  \widetilde{c} = c + \nu,\quad \nu \sim \mathcal{U}\{0,1,\dots,B-1\},
  \label{eq:noisemechanism}
\end{equation}
where $B$ is a power-of-two noise bound committed on-chain when the dataset
first opts into noise, $c$ is the true encrypted aggregate, and $\nu$ is
sampled by the FHE coprocessor via a width-matched
\texttt{FHE.randEuintT(B)} inside the \texttt{injectQueryNoise} call, with
$T{=}64$ for T3/G1 and $T{=}32$ for T4/T5 \emph{(e.g., a dataset opting in with
$B = 8$ releases counts perturbed by integer noise in \{0, ..., 7\} with
expectation $(B{-}1)/2 = 3.5$)}.  After injection, the released handle
$\widetilde{c}$ is the only value the requester ever decrypts; the true $c$
and the realized noise $\nu$ are both unrecoverable from on-chain state.  The
mechanism is not a differential-privacy guarantee: because the noise is one-sided
and $\Pr[\nu=0]=1/B$, deployments must combine it with per-key rate limits,
query deduplication or cached noisy answers, or a composition-aware mechanism if
formal privacy is required.
Figure~\ref{fig:noise} shows the lifecycle position of the injection.  In the
current implementation, query-noise injection is implemented for the genotype
tier lineage and the G1 multi-filter contract; single-dimension sex, age, and
phenotype filter contracts currently return exact encrypted counts and require
the same hardening as future work.

\subsection{Trust Model}

The coordinator triggers \texttt{injectQueryNoise(queryId)} with no noise
parameter; the contract internally calls the width-matched \texttt{FHE.randEuintT}.
The realized $\nu$ cannot be observed or influenced by the coordinator before
the block is mined.  Coordinator-chosen zero-noise injection --- the central weakness
of any off-chain noise scheme where a malicious coordinator could choose
$\nu = 0$ --- is therefore prevented by construction.  Natural
zero-noise draws still occur with probability $1/B$.  The remaining
failure mode is liveness rather than confidentiality: a withheld
\texttt{injectQueryNoise} call leaves the query stuck, but the cohort is still
protected.  Two further hardening guards are in place: $B$ must be a positive
power of two (a coprocessor requirement that also enforces a discrete
log-scale parameter), and $B$ becomes immutable after the first noise-enable
call (an implementation hardening described in Section~\ref{sec:security}).

\subsection{Calibration}

The discrete uniform noise on $\{0,1,\ldots,B-1\}$ has expected value
$(B-1)/2$, so released counts carry an upward bias.

\emph{Repeated-query collapse.}  Because the noise is one-sided and zero is in
the support, an adversary who issues the same query $k$ times and keeps the
minimum recovers the true count whenever any draw is zero.  The probability
of at least one zero draw across $k$ identical queries is
\begin{equation}
  \Pr[\nu_{\min}=0 \mid k] \;=\; 1 - \left(1 - \tfrac{1}{B}\right)^{k},
  \label{eq:nuzero}
\end{equation}
which already exceeds $\tfrac{1}{2}$ at $k \approx B \ln 2 \approx 0.69\,B$
repeated queries.  Table~\ref{tab:noise-collapse} tabulates a few practical
points: even at $B{=}16$ a 20-query budget already gives a $72\%$ chance of
collapsing the protection.  Bounded noise therefore must be combined with
per-key rate limits, cached noisy answers for duplicate query keys, or a
composition-aware mechanism if formal repeated-query privacy is required;
on its own it is an anti-probing cost increment, not a closed channel.
Practical operating choices for $B$ are listed in
Table~\ref{tab:noise-calibration}; for multi-filter (G1) queries the
minimum subgroup count may be small even when marginal counts are large,
so $B \geq 8$ is recommended there.

\begin{table}[t]
  \centering
  \caption{\textbf{Min-attack collapse probability.}
  $\Pr[\nu_{\min}=0\mid k]=1-(1-1/B)^k$ as a function of bound $B$ and
  number of identical repeated queries $k$.  Rate-limit policy must keep
  $k$ small enough that this probability remains acceptable.}
  \label{tab:noise-collapse}
  \small
  \begin{tabularx}{\columnwidth}{lYYYY}
    \toprule
    $B$ & $k{=}1$ & $k{=}3$ & $k{=}10$ & $k{=}20$ \\
    \midrule
    $4$  & $25.0\%$ & $57.8\%$ & $94.4\%$ & $99.7\%$ \\
    $8$  & $12.5\%$ & $33.0\%$ & $73.7\%$ & $93.1\%$ \\
    $16$ & $6.25\%$ & $17.6\%$ & $47.4\%$ & $72.3\%$ \\
    $32$ & $3.13\%$ & $9.1\%$  & $27.2\%$ & $47.0\%$ \\
    \bottomrule
  \end{tabularx}
\end{table}

\begin{table}[t]
  \centering
  \caption{\textbf{Practical noise-bound choices.}  Heuristic operating
  points; for G1 queries use $B \geq 8$ because $4$-way subgroup counts can
  be small even when marginals are large.}
  \label{tab:noise-calibration}
  \small
  \begin{tabularx}{\columnwidth}{lYY}
    \toprule
    $B$ & $\mathbb{E}[\nu]$ & Heuristic use \\
    \midrule
    $4$ & $1.5$ & Very rare variants ($c \leq 5$) \\
    $8$ & $3.5$ & Rare variants; G1 (subgroup $\leq 10$) \\
    $16$ & $7.5$ & General use; G1 (subgroup $\leq 50$) \\
    $32$ & $15.5$ & High-count markers ($c > 100$) \\
    \bottomrule
  \end{tabularx}
\end{table}

\subsection{Cost}

The injection adds one width-matched \texttt{FHE.randEuintT} and one
\texttt{FHE.add} per query.  On the T3/G1 \texttt{euint64} path this costs
$\approx 215{,}000$ EVM gas and $\approx 162{,}000$ HCU.
Relative to the per-query scan cost, the overhead drops from $10.8\%$ at
$N{=}31$ to $<\!1\%$ at $N{=}500$.  The gas overhead is small relative to the
scan cost, but the privacy value still depends on rate-limit and deduplication
policy.

\section{Security Model}
\label{sec:security}

\subsection{Threat Model}

We assume eight classes of adversary, drawn from the project's design
document.  A \emph{curious requester} is authorized but tries to extract more
than the aggregate they were promised; the mitigation is that no
per-contributor handle is ever granted via \texttt{FHE.allow} to anyone but the aggregator.
\emph{Mutually distrustful contributors} share the compute layer; per-contributor
handles never leave the contract's own ACL.  An \emph{external chain observer}
sees all transactions and ciphertext handles but cannot decrypt without the
KMS-held key.  A \emph{rate-limit attacker} attempts to count-difference
across many queries; the per-(dataset, requester) windowed quota and, where
implemented, the bounded noise of Section~\ref{sec:noise} raise the cost of the
attack, but do not provide formal privacy under unrestricted repetition.  Note
that bounded query-noise is only applied on $4$ of the $13$ implemented query
paths (genotype T3/T4/T5 and the G1 multi-filter); the $9$ single-axis sex,
age, and phenotype filter contracts return exact encrypted counts and rely
entirely on rate limits and governed requester approval against this
adversary.  A \emph{coordinator} controls policy and approvals but cannot read
encrypted state or substitute noise values.  A \emph{storage griefer} can
create queries and never finalize them; \texttt{cancelStaleQuery} reclaims them
after the time-to-live (TTL) window.  \emph{Malicious contributors} can submit
garbage encrypted counts; upload integrity is out of scope for v1
(Section~\ref{sec:limitations}).  Finally, the \emph{KMS / decryption gateway}
is an explicit trust assumption: result decryption is delegated to an off-chain
key-management service that holds the threshold-shared decryption key (the
Zama threshold KMS in the reference fhEVM deployment).  This service is
trusted not to release the encrypted aggregate to any address other than the
one named in \texttt{FHE.allow(handle, addr)}; a malicious or compromised KMS
can expose result handles, and bioETH-Beacon does not eliminate this trust
assumption.

\begin{figure}[t]
  \centering
  \includegraphics[width=\linewidth]{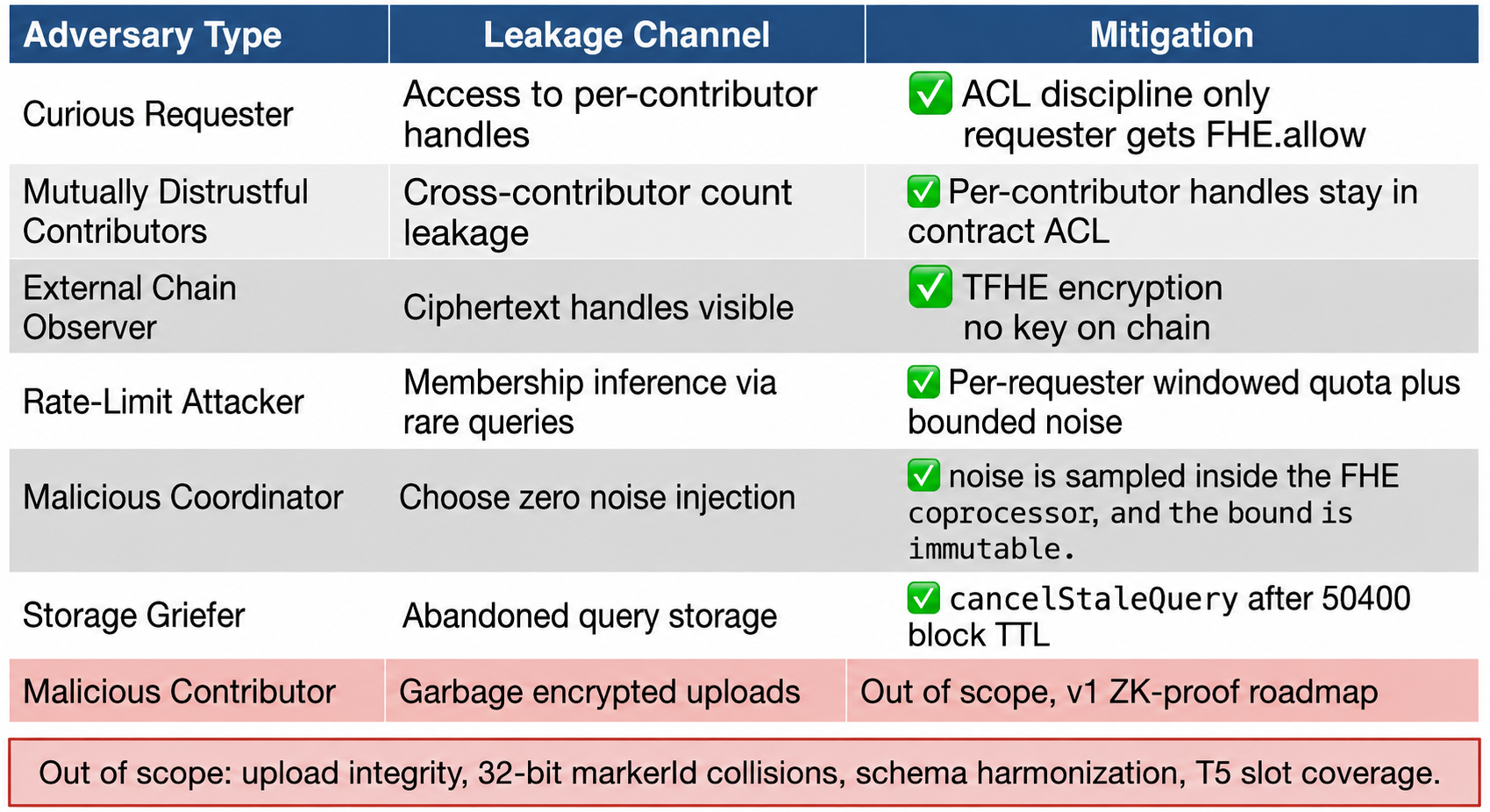}
  \caption{\textbf{Threat model (seven on-chain adversaries).}  Each
  on-chain leakage channel is mapped to its on-chain mitigation; out-of-scope
  channels (upload integrity, population-scale collisions, schema
  harmonization, KMS compromise) are documented as limitations rather than
  silently glossed.  KMS compromise --- the eighth adversary in the threat
  model --- is an off-chain trust assumption (see Section~\ref{sec:security})
  and is not represented in this diagram.}
  \label{fig:threat_model}
\end{figure}

What the system explicitly does \emph{not} protect: (i) site-level
participation visibility through upload events; (ii) marker-ID collisions at
population scale ($> 2^{16}$ distinct markers); (iii) schema harmonization
errors across contributors that produce silently incorrect aggregates; (iv) T5
slot-index visibility revealing which markers each contributor tracks; (v)
arbitrary $N$-way conjunctive queries beyond the four-predicate G1 model; and
(vi) KMS compromise or collusion at the off-chain decryption gateway.
Figure~\ref{fig:threat_model} maps the seven on-chain leakage channels to
their on-chain mitigations.  The eighth adversary class introduced above ---
KMS / decryption-gateway compromise --- is an off-chain trust assumption
rather than an on-chain leakage channel and is therefore documented in the
text and the out-of-scope list (item vi) rather than in the figure.

\subsection{Invariants}

Eleven invariants are stated explicitly in the codebase's implementation
security checklist and enforced contract by contract.  Each is short enough to state directly.

\begin{invariant}\label{inv:nopt}
For T3/T4, no plaintext genomic marker IDs, filter values, or counts appear in
contract storage or events.  For T5, counts remain encrypted but marker-slot
indices and presence are intentionally public.
\end{invariant}
\begin{invariant}\label{inv:reqpriv}
Query results remain requester-private; aggregate counts are never made
publicly decryptable.
\end{invariant}
\begin{invariant}\label{inv:fullscan}
Query execution scans the full configured entry geometry; the inner loop has
no encrypted-conditional early exit.
\end{invariant}
\begin{invariant}\label{inv:authcreate}
\texttt{createQuery} enforces registry authorization, retirement status, and
local rate limits before any FHE state is created.
\end{invariant}
\begin{invariant}\label{inv:authupload}
\texttt{appendContributorChunk} requires explicit registry approval of the
caller before reaching the compute layer.
\end{invariant}
\begin{invariant}\label{inv:allowthis}
Every encrypted handle persisted in storage calls \texttt{FHE.allowThis}.
\end{invariant}
\begin{invariant}\label{inv:finalize}
\texttt{finalizeQuery} grants result access only to the requester.
\end{invariant}
\begin{invariant}\label{inv:rosterlock}
The contributor roster is locked before dataset finalization.
\end{invariant}
\begin{invariant}\label{inv:epoch}
Dataset epochs are append-free after finalization; new data requires a new
shell registration.
\end{invariant}
\begin{invariant}\label{inv:chunksize}
\texttt{queryChunkSize} does not exceed the empirically validated safe value.
The current profile/benchmark value is $29$ for the \texttt{euint64} accumulator
path; deployment manifests use $23$ as a $\approx\!80\%$ safety margin unless a
fresh HCU probe justifies a higher value.  The analytic add-chain ceiling is
$37$, but $37$ fails in the current probe; values in $30$--$36$ have not been
tested.
\end{invariant}
\begin{invariant}\label{inv:mincontrib}
Datasets with \texttt{minContributors}~$> 0$ cannot be finalized until that many
distinct institutions have uploaded.
\end{invariant}

Invariants~\ref{inv:nopt}--\ref{inv:epoch} and~\ref{inv:mincontrib} are enforced
in contract code; Invariant~\ref{inv:chunksize} is governance-enforced via the
pre-deployment advisor and surfaces on-chain as the failure mode
\texttt{HCUTransactionDepthLimitExceeded} when violated at runtime.

\section{Empirical Evaluation}
\label{sec:eval}

\subsection{Methodology}

\textbf{All reported gas figures in this paper are local Hardhat EVM-gas, not
live-network total cost.}  Hardhat does not include the HCU adder, which on a
live fhEVM coprocessor is comparable to or larger than the EVM floor; live
total per-query cost is therefore expected to be materially higher than the
figures reported here, while tier-relative deltas (T3/T4/T5) and asymptotic
shape (linear in $N$, flat in $M$ for T5) are expected to carry over.  All
measurements come from Hardhat tests, which provide reproducible local EVM-gas
and sequential-depth measurements for the evaluated contracts; live network
execution may impose additional accounting and deployment-specific
configuration.  We therefore report EVM gas \emph{measured} and HCU
\emph{analytic}, with the HCU model calibrated against the per-operation costs
published in Zama's HCU table \citep{zamafhevmdocs} and confirmed against the
empirical depth probe.  The
synthetic count generator draws marker presence from a Bernoulli model
across hospitals at densities $D\in\{0.2\%, 1\%, 5\%, 10\%\}$ with
negative-binomial counts; marker panels are derived deterministically from the
PGS Catalog \citep{lambert2021pgscatalog}.  Five panel sizes are reported:
P$16$ (the hand-curated worked example), P$100$, P$1$K, P$10$K, P$100$K.

\subsection{Worked Consortium Example}
\label{sec:worked-example}

To make the benchmark concrete, the P$16$ panel is instantiated as a four-hospital
consortium with a shared dictionary of $16$ canonical single-nucleotide variants
and $31$ sparse non-zero encrypted uploads.  \emph{Throughout this worked
example we illustrate the data-flow at} \texttt{uploadChunkSize=4} \emph{and}
\texttt{queryChunkSize=5} \emph{for narrative clarity; all measured gas
reported in this paper, including in this subsection, uses the operational
defaults} \texttt{uploadChunkSize=16} \emph{and} \texttt{queryChunkSize=29}
\emph{(the largest tested passing value; deployment manifests use $23$ as a
safety margin).}  The example uses genome build \texttt{GRCh38}, dictionary
version \texttt{marker-matrix-v2026-05}, and the normalization rule
\texttt{SNV\_CANON\_V1}; each hospital keeps its dense local row off-chain and
uploads only encrypted non-zero marker--count pairs.
Table~\ref{tab:worked-upload} summarizes the sparse upload footprint: all four
institutions share the same dictionary, but each submits only its non-zero
marker counts.  This mirrors the deployed contract path, where upload privacy
is provided by encrypted handles rather than by hiding transaction counts.
With the illustrative \texttt{queryChunkSize=5} a full T3 query scans all
$31$ encrypted entries in seven chunks; at the operational
\texttt{queryChunkSize=29} the same workload completes in a single chunk.
Only the authorized requester can retrieve the plaintext final aggregate, through the off-chain KMS-mediated decryption path.

\begin{table}[t]
  \centering
  \caption{\textbf{Worked P$16$ upload footprint.}  Dense hospital rows remain
off-chain; only non-zero encrypted marker-count entries are uploaded.}
  \label{tab:worked-upload}
  \small
  \begin{tabularx}{\columnwidth}{Yrr}
    \toprule
    Institution & Entries & Upload chunks \\
    \midrule
    North Regional Hospital & $8$ & $2$ \\
    East Oncology Centre & $8$ & $2$ \\
    River Children's Genomics Unit & $8$ & $2$ \\
    Metro Precision Medicine Lab & $7$ & $2$ \\
    \bottomrule
  \end{tabularx}
\end{table}

\begin{table}[t]
  \centering
  \caption{\textbf{Worked P$16$ consortium queries.}  Expected counts are computed
  from the four-hospital marker matrix; decrypted counts are produced by the Hardhat
  end-to-end validation path.}
  \label{tab:worked-example}
  \small
  \begin{tabularx}{\columnwidth}{lYrr}
    \toprule
    Query & Canonical variant & Expected & Decrypted \\
    \midrule
    EG-1 & \texttt{chr7:117199644:C>T} & $43$ & $43$ \\
    EG-2 & \texttt{chrX:153296891:A>G} & $26$ & $26$ \\
    EG-3 & \texttt{chr22:19952036:C>T} & $14$ & $14$ \\
    \bottomrule
  \end{tabularx}
\end{table}

\subsection{Per-Step EVM Gas}

Table~\ref{tab:gas-baseline} summarizes EVM gas for the T3 reference path at
three synthetic working points.  Upload cost is dominated by
\texttt{FHE.fromExternal} verification and SSTORE of two encrypted handles per
entry: per-entry upload costs $\sim\!205$k gas across all measured $N$, and
gas scales linearly with entry count.  Query cost is the
$\texttt{eq}\!\to\!\texttt{select}\!\to\!\texttt{add}$ chain accumulated chunk-by-chunk;
the per-chunk receipt absorbs $29$ scan iterations, each containing the
\texttt{eq}$\to$\texttt{select}$\to$\texttt{add} FHE chain, plus the
\texttt{SSTORE} of the updated handle.  In local Hardhat tests, the dominant cost is EVM
gas; on a live network the HCU adder is comparable to or larger than the
EVM floor (Section~\ref{sec:eval-hcu}).

\begin{table}[t]
  \centering
  \caption{\textbf{Per-step EVM gas (T3 reference path).}
  Configuration: \texttt{uploadChunkSize}~$=16$, \texttt{queryChunkSize}~$=29$
  (largest tested passing value; deployment manifests use $23$), \texttt{euint32}
  markers, \texttt{euint64} counts.  \texttt{createQuery} is structurally
  independent of $N$; the $\le\!12$-gas spread across columns is measurement
  jitter.  Source:
  \texttt{data/reports/gas-profile-hardhat.json}.}
  \label{tab:gas-baseline}
  \small
  \resizebox{\columnwidth}{!}{%
  \begin{tabular}{lrrr}
    \toprule
    Step & $N{=}31$ & $N{=}100$ & $N{=}500$ \\
    \midrule
    \texttt{createDatasetShell} & $147{,}577$ & $147{,}601$ & $147{,}613$ \\
    \texttt{createDatasetStorage} & $82{,}881$ & $82{,}881$ & $82{,}881$ \\
    \texttt{approveContributor} & $73{,}061$ & $73{,}061$ & $73{,}061$ \\
    \texttt{appendContributorChunk}$\times K$ & $6{,}370{,}566$ & $20{,}503{,}156$ & $102{,}387{,}074$ \\
    \texttt{lockContributors} & $32{,}447$ & $32{,}447$ & $32{,}447$ \\
    \texttt{finalizeDataset} & $44{,}526$ & $44{,}526$ & $44{,}526$ \\
    \texttt{grantQueryAccess} & $50{,}443$ & $50{,}443$ & $50{,}443$ \\
    \texttt{createQuery} & $317{,}939$ & $317{,}927$ & $317{,}939$ \\
    \texttt{processQueryChunk}$\times C$ & $1{,}608{,}175$ & $4{,}942{,}565$ & $24{,}517{,}886$ \\
    \texttt{finalizeQuery} & $76{,}570$ & $76{,}570$ & $76{,}570$ \\
    \bottomrule
  \end{tabular}%
  }
\end{table}

\subsection{Tier Comparison at \texorpdfstring{$N{=}500$}{N=500}}

Table~\ref{tab:tier-comparison} compares the three tiers at the common working
point $N{=}500$.  T3$\to$T4 leaves EVM gas effectively unchanged in local Hardhat tests
($-0.08\%$ upload, $-0.20\%$ query) but reduces HCU per entry by $24\%$ on
the analytic model (Section~\ref{sec:eval-hcu}).  T3$\to$T5 yields the
fixed-$M{=}20$ headline result: query EVM gas drops by $94\%$ (from
$24{,}912{,}395$ to $1{,}449{,}177$) and upload EVM gas by $53\%$ (from $102{,}469{,}955$ to
$47{,}739{,}867$).  The one-time \texttt{createDatasetStorage} cost rises
from $82{,}881$ to $937{,}602$ gas to initialise the $M{=}20$ encrypted
slots; the per-query saving is substantial, but only worthwhile when queries
outnumber dataset epochs and the trust model permits public slot indices.

\begin{table}[t]
  \centering
  \caption{\textbf{Tier comparison at $N{=}500$.}  T5 figures use $M{=}20$
  slots.  \emph{Upload gas} aggregates \texttt{createDatasetStorage} +
  \texttt{appendContributorChunk}; \emph{Query gas} aggregates
  \texttt{createQuery} + \texttt{processQueryChunk} + \texttt{finalizeQuery}.
  Source: fixed-$M{=}20$ optimization/profile run summarized in
  \texttt{reports/optimization-report.md}.}
  \label{tab:tier-comparison}
  \small
  \resizebox{\columnwidth}{!}{%
  \begin{tabular}{lrrr}
    \toprule
    & T3 & T4 & T5 ($M{=}20$) \\
    \midrule
    Upload gas & $102{,}469{,}955$ & $102{,}393{,}043$ & $47{,}739{,}867$ \\
    Query gas & $24{,}912{,}395$ & $24{,}861{,}752$ & $1{,}449{,}177$ \\
    Query tx count & $18$ & $18$ & $1$ \\
    HCU unit & $277$k/entry & $211$k/entry & $211$k/slot \\
    $\Delta_{T3}$ upload & --- & $-0.08\%$ & $-53.4\%$ \\
    $\Delta_{T3}$ query & --- & $-0.20\%$ & $-94.2\%$ \\
    \bottomrule
  \end{tabular}%
  }
\end{table}

\begin{figure}[t]
  \centering
  \includegraphics[width=\linewidth]{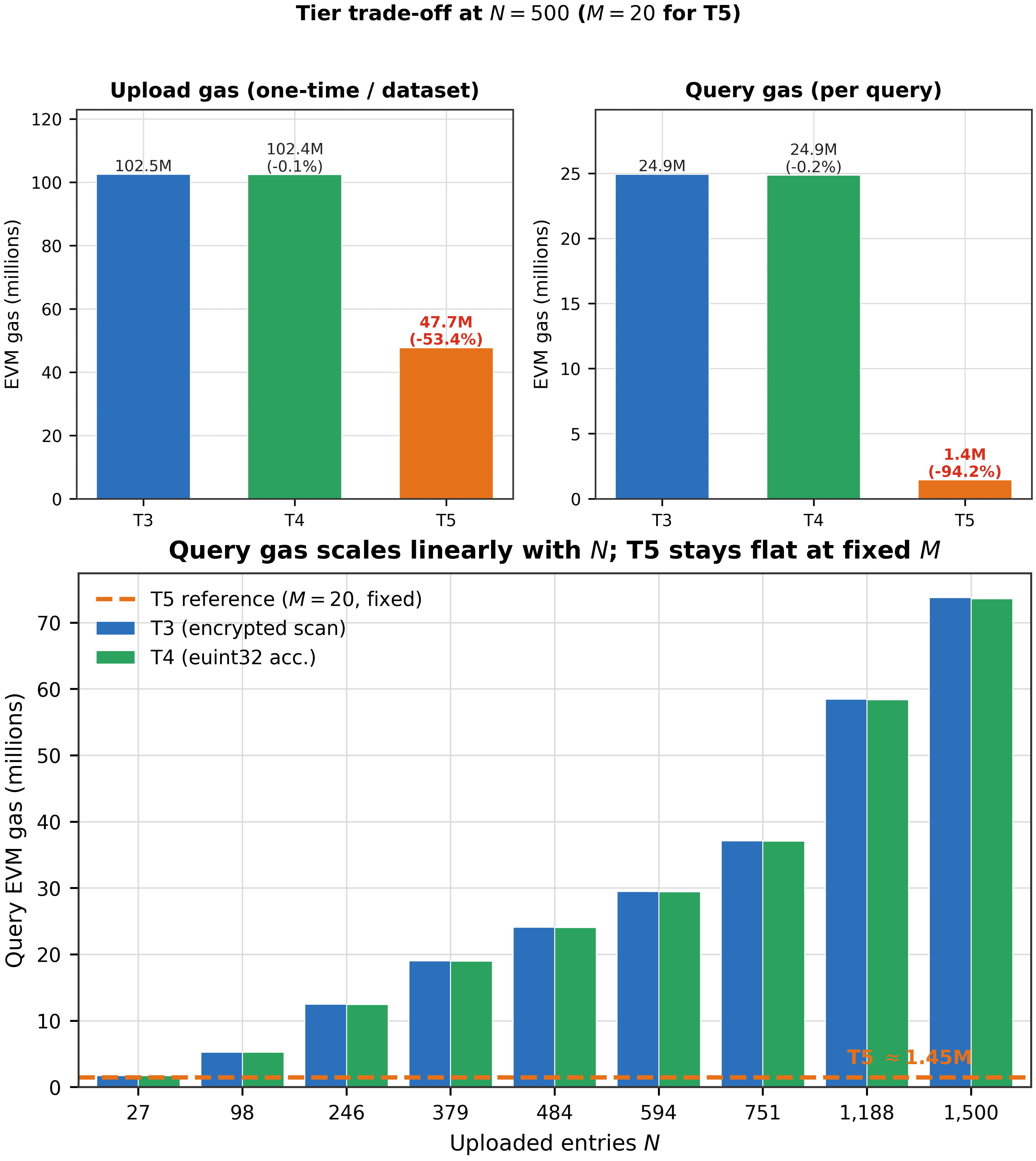}
  \caption{\textbf{Tier-level gas tradeoffs and scaling behavior (local
  Hardhat EVM-gas).}
  (A) Fixed-\(N{=}500, M{=}20\) comparison of upload and query \emph{local}
  EVM gas across T3, T4, and T5.  T4 is EVM-gas-equivalent to T3 in local
  Hardhat tests, while T5 reduces query gas by $94\%$ and upload gas by
  $53\%$ relative to T3.  (B) Local EVM query gas as a function of uploaded
  entries \(N\).  T3 and T4 scale linearly with \(N\), whereas T5 remains
  approximately flat at fixed \(M{=}20\) until \(M\) itself grows.  Live
  per-query cost includes an HCU adder not reflected here.}
  \label{fig:tier-tradeoff-scaling}
\end{figure}

\subsection{Scaling to PGS Catalog Panels}
\label{sec:eval-scaling}

Table~\ref{tab:scaling} reports the cross-product of panel size $\times$ hospital
count $\times$ density $\times$ tier on the synthetic benchmarks.  The key
findings: (i) per-entry upload gas stays in the $204$k-$205$k band across
\emph{every} configuration, confirming linear scaling through P$100$K; (ii)
P$10$K with $16$ hospitals at $1\%$ density ($N{=}1{,}500$) completes
correctness verification in $\sim\!6$s of wall time in local Hardhat tests and consumes
$73.8$M query gas across $54$ chunks; (iii) in the current 32-bit on-chain
marker-ID truncation used by the benchmark path, P$100$K shows exactly one
observed collision among $100{,}000$ benchmark markers ($0.001\%$ of IDs),
illustrating why production manifests must reject colliding dictionaries or
migrate to wider marker types such as \texttt{euint128}; (iv) T5 crosses over with T3 between $N{=}246$ and $N{=}484$ on the
P$1$K panel --- below the crossover, T3 is cheaper; above it, T5 wins by a
widening margin.  Figure~\ref{fig:tier-tradeoff-scaling}B renders the
cross-tier comparison as a function of $N$.

\begin{table*}[t]
  \centering
  \caption{\textbf{Measured scaling across PGS Catalog panels.}  Tier T3 with
  \texttt{queryChunkSize}~$=29$.  Source: measured Hardhat panel benchmark run
  summarized in \texttt{reports/benchmark-scaling.md} from
  \texttt{data/reports/panel-benchmark.json} and
  \texttt{data/reports/panel\_benchmark.csv}.}
  \label{tab:scaling}
  \small
  \begin{tabular}{lrrrrrrr}
    \toprule
    Panel & Markers & Hosp. & $N$ & Upload gas & Query gas & Query tx & Wall time \\
    \midrule
    P$100$, $D{=}5\%$ & $100$ & $4$ & $27$ & $5{,}361{,}001$ & $1{,}742{,}352$ & $3$ & $161$ms \\
    P$100$, $D{=}10\%$ & $100$ & $8$ & $98$ & $19{,}836{,}557$ & $5{,}269{,}524$ & $6$ & $420$ms \\
    P$1$K, $D{=}5\%$ & $1{,}000$ & $4$ & $246$ & $50{,}331{,}922$ & $12{,}494{,}011$ & $11$ & $958$ms \\
    P$10$K, $D{=}1\%$ & $10{,}000$ & $4$ & $379$ & $77{,}633{,}597$ & $19{,}038{,}003$ & $16$ & $1{,}483$ms \\
    P$1$K, $D{=}5\%$ & $1{,}000$ & $8$ & $484$ & $98{,}805{,}844$ & $24{,}107{,}664$ & $19$ & $1{,}918$ms \\
    P$100$K, $D{=}0.2\%$ & $100{,}000$ & $4$ & $594$ & $121{,}585{,}680$ & $29{,}506{,}183$ & $23$ & $2{,}323$ms \\
    P$10$K, $D{=}1\%$ & $10{,}000$ & $8$ & $751$ & $153{,}712{,}561$ & $37{,}138{,}985$ & $28$ & $2{,}953$ms \\
    P$100$K, $D{=}0.2\%$ & $100{,}000$ & $8$ & $1{,}188$ & $243{,}137{,}208$ & $58{,}494{,}919$ & $43$ & $4{,}855$ms \\
    P$10$K, $D{=}1\%$ & $10{,}000$ & $16$ & $1{,}500$ & $306{,}732{,}738$ & $73{,}771{,}830$ & $54$ & $6{,}046$ms \\
    \bottomrule
  \end{tabular}
\end{table*}

\begin{figure}[t]
  \centering
  \includegraphics[width=\linewidth]{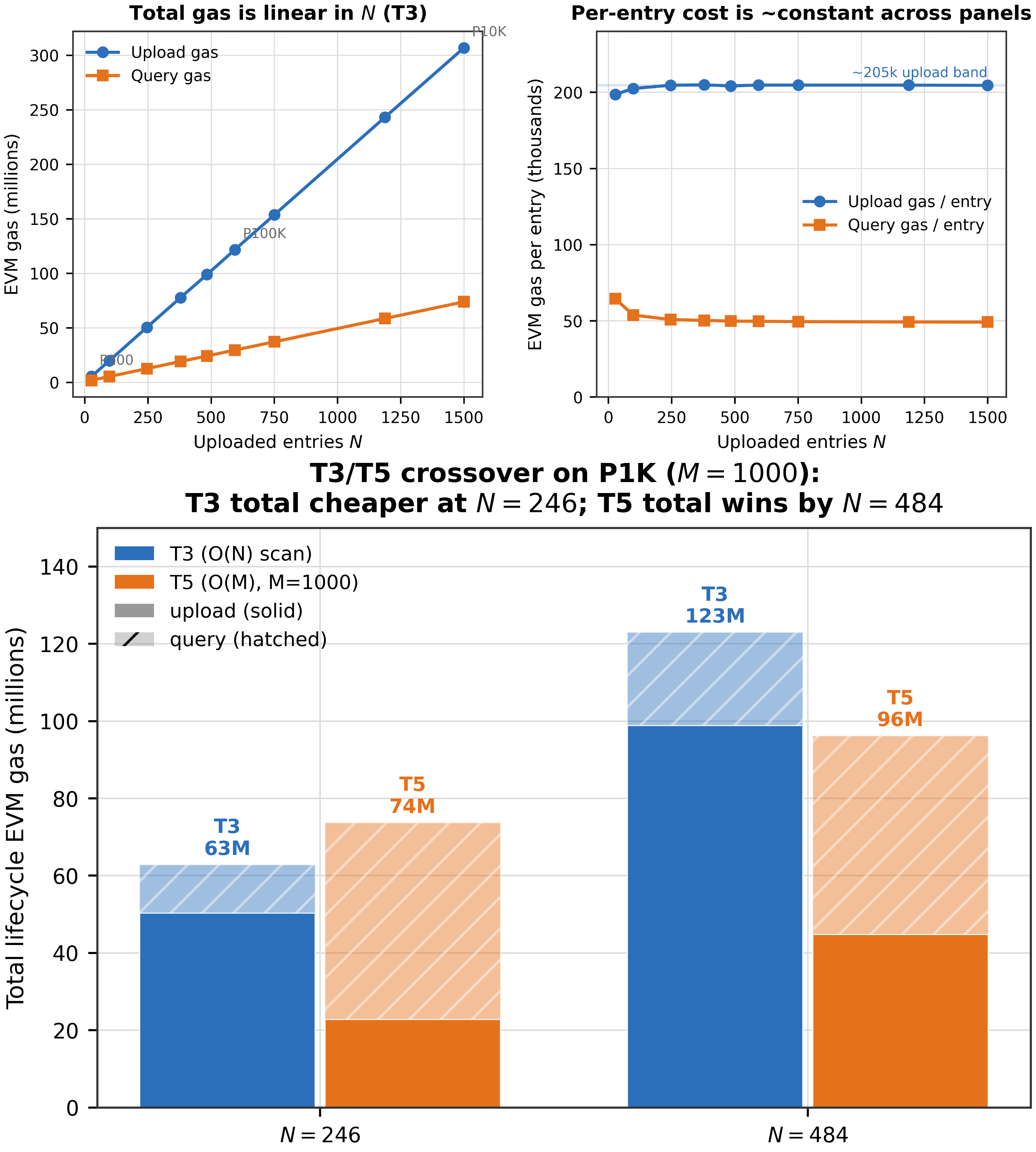}
  \caption{\textbf{Scaling behavior and lifecycle crossover across benchmark
  panels (local Hardhat EVM-gas).}
  (A) Total local EVM query gas grows linearly with uploaded entries \(N\)
  across PGS Catalog-derived panel sizes.  (B) Per-entry upload gas remains
  approximately constant at \(\approx 205\)k across configurations, confirming
  that upload cost depends mainly on the number of non-zero entries rather
  than panel size.  (C) T3 vs. T5 total lifecycle gas crossover on P1K.
  T3 total gas is lower below \(N \approx 246\), while T5 wins above
  \(N \approx 484\) because its lower upload cost offsets the fixed-slot
  query model.  Crossover thresholds shift with live HCU pricing.}
  \label{fig:scaling-lifecycle-crossover}
\end{figure}

\subsection{HCU Cost Model}
\label{sec:eval-hcu}

The per-entry HCU adder is dominated by the accumulator step.  For T3:
$\texttt{FHE.eq}(\texttt{euint32}) = 60$k, $\texttt{FHE.select}(\texttt{euint64}) = 55$k,
$\texttt{FHE.add}(\texttt{euint64}) = 162$k, totalling $277$k HCU per scanned
entry.  T4's narrower accumulator reduces this to $211$k ($-24\%$).  T5 evaluates
the same kernel but only $M$ times regardless of $N$, so per-entry HCU is
reported as $211$k per slot and amortized over $M$ slots per epoch.  The \texttt{FHE.add} operation contributes $162$k to the global HCU budget but
$133$k to the sequential-depth budget in Zama's two-limit model; the depth
budget binds first.  The analytic add-chain ceiling is
$\lfloor 5\!\times\!10^6 / 133{,}000\rfloor = 37$, but the current probe fails
at \texttt{queryChunkSize}~$=37$; the largest tested passing value is $29$,
which is used by every measurement in this section.  Deployment manifests use
$23$ as a $\approx\!80\%$ safety margin unless a fresh HCU probe justifies
higher.  Values in the range $30$--$36$ remain untested, so the true empirical
ceiling lies in that interval; the gap between the analytic $37$ and the
empirical $\le 36$ reflects fixed per-chunk overhead in the depth accounting
beyond the bare add-chain estimate.

\subsection{Noise-Injection Overhead}

The bounded-noise injection (Section~\ref{sec:noise}) adds
$\approx 215{,}000$ EVM gas and $\approx 162{,}000$ HCU per query.  As a
fraction of total query cost: $10.8\%$ at $N{=}31$, $4.0\%$ at $N{=}100$,
and $0.9\%$ at $N{=}500$.  At realistic consortium scales the gas overhead of noise injection is small
relative to the scan cost; its privacy value still depends on rate-limit and
deduplication policy.

\subsection{Multi-filter (G1)}

The conjunctive G1 contract performs $4\!\times\!\texttt{FHE.eq} + 3\!\times\!\texttt{FHE.and}$
per entry on top of the base \texttt{select}/\texttt{add} pair.  Mock-measured EVM gas
is essentially identical to T3 ($\Delta < 0.001\%$ per upload entry; $0\%$ per
query chunk at $N{=}500$); the additional cost is entirely in the HCU
adder, which rises from $277$k to $\sim\!472$k per entry ($+70\%$).  Two
constraints converge at the multi-filter chunk ceiling: sequential depth
limits chunks to $\sim\!37$ entries, and the global HCU budget limits them
to $\sim\!42$.  Sequential depth binds first ($37 < 42$); we therefore
adopt $23$ as the conservative G1 deployment-manifest ceiling and use $29$
(the largest tested passing value, inherited from the T3 profile) for the
measurements reported here.  The same pattern is summarized by the
filter-family benchmarks.

\subsection{Correctness and Tests}

Correctness is verified end-to-end at three scales.  The hand-curated
$N{=}31$ worked example decrypts to expected
counts on three reference queries ($43$, $26$, $14$).  The scaling test
suite verifies $\mathit{decrypted} = \mathit{expected}$ across seven
synthetic configurations from P$100$ to P$10$K, including queries against
guaranteed-zero markers.  The Hardhat test suite covers unit, integration,
scaling, single-filter (sex/age/phenotype $\times$ T3/T4/T5), multi-filter
(G1), and noise-injection cases.  In the version of the suite reported here,
$181$ tests pass and $2$ are pending: both pending cases are live-coprocessor
T4 chunk-ceiling probes that require a funded Sepolia deployment and could not
be executed in local Hardhat.  The graphical abstract additionally cites
``$0$ critical findings'' from the internal security review documented in
\texttt{reports/security-review.md}, which audits the eleven invariants of
Section~\ref{sec:security} against the implemented contract suite; that
internal review is not a substitute for independent third-party audit.

\section{Discussion and Limitations}
\label{sec:limitations}

\subsection{When to Pick Which Tier}

In the fixed-$M{=}20$ benchmark, T3 remains the default privacy-preserving
choice below roughly $500$ uploaded entries when full marker-presence privacy is
required; operators should recompute the crossover for their $M$, query volume,
and live HCU pricing.  T3 pays a modest gas cost in exchange for full encryption
of marker IDs, filter values, and counts.  T4 is appropriate when the maximum
true aggregate plus optional noise fits in \texttt{uint32} and the operator
targets a live coprocessor where the $24\%$ HCU saving is visible.  T5 is the
right choice when (i) the set of markers each contributor tracks is not sensitive,
(ii) queries dominate uploads ($Q/U \gg 1$), and (iii) the panel is dense
enough that $M \ll N$.  At $N=500$ and $M=20$, T5 cuts query gas by
$94\%$; at $M\!\approx\!N$ the advantage vanishes.  Two distinct comparisons
appear in the evaluation and should not be conflated: the $94\%$ figure is
\emph{query-gas only} at fixed $M{=}20$, whereas the $N\!\approx\!246$--$484$
crossover reported in Section~\ref{sec:eval-scaling} is \emph{total lifecycle
gas} (upload + query) and shifts with $M$, the query/upload volume ratio,
and live HCU pricing.

\subsection{What the System Does Not Cover}

\textbf{Upload integrity.}  A malicious contributor can submit any
encrypted value.  FHE provides confidentiality, not authenticity; the
contract cannot verify that the submitted counts match the contributor's
real electronic health record (EHR).  ZK proofs binding uploads to off-chain attestations are
roadmap work.

\textbf{Marker collisions at population scale.}  The \texttt{euint32}
encoding has non-negligible birthday risk as dictionaries grow: for a uniform
$32$-bit prefix, the probability of at least one collision is about $1.2\%$ at
$10{,}000$ markers and about $39\%$ at $65{,}536$ markers, reaching effective
certainty long before $10^7$.  Three operating regimes therefore apply.  Below
$\sim\!10^4$ markers, the consortium manifest must perform an off-chain
collision check and reject any colliding dictionary outright (\emph{deterministic
reject-and-remap}).  Between $\sim\!10^4$ and $\sim\!6.5\!\times\!10^4$ markers,
reject-and-remap remains feasible but increasingly costly, and an explicit
collision table mapping colliding canonical variants to disambiguated
\texttt{markerId} values is required.  Above $\sim\!10^5$ markers,
reject-and-remap becomes impractical and migration to the wider
\texttt{euint128} encoding is mandatory (at $\sim\!3\!\times$ HCU per
\texttt{eq}).  A $10^7$-marker threshold describes only a
catastrophic-collision regime; the operating threshold for the
\texttt{euint128} migration is closer to $10^5$.

\textbf{Schema harmonization.}  If contributors apply inconsistent variant
normalizations, the aggregate can be silently incorrect.  The dictionary URI and
commitment in \texttt{BeaconRegistry} provide an off-chain integrity
anchor but the contract itself cannot enforce normalization.

\textbf{Event-timing leakage.}  \texttt{QueryCreated} and
\texttt{QueryFinalized} reveal the requester address, dataset, query timing,
chunk progress, and finalization timing, without exposing the marker or the
count.  This metadata leakage is an accepted gap.

\textbf{Cross-contract differencing on the same cohort.}  If a cohort is
exposed simultaneously through a single-filter T3 and the G1
multi-filter, an attacker authorized on both can reconstruct
multi-dimensional marginals without exhausting either rate-limit window.
Operational guidance is to restrict G1 access to a strictly smaller
requester set than T3, or to share rate-limit accounting across related
datasets in a registry extension.

\subsection{Roadmap}

Documented future items include encrypted indexed lookup (an
$O(\log M)$ FHE binary trie replacing the linear scan), dataset
versioning with explicit deprecation, time-locked or multisig
coordinator governance, and an \texttt{euint128} migration for
population-scale catalogs.  None of
these are required for the current evaluation; each is a separable
extension that does not invalidate the present design.

\section{Related Work}
\label{sec:related}

\textbf{The Beacon protocol and its privacy.}
Fiume \emph{et al.}\ \citep{fiume2019federated} formalized the Beacon Network
as a federated genomic discovery layer with existence and optional quantitative
variant-level responses; Rambla \emph{et al.}\ \citep{rambla2022beacon}
introduced Beacon v2 as a broader biomedical-discovery model with structured
result sets and richer filtering terms.
Shringarpure and Bustamante \citep{shringarpure2015privacy} first
demonstrated membership inference against Beacons, and Raisaro
\emph{et al.}\ \citep{raisaro2017addressing} proposed budget-based defenses.
Both threat models assume a benign Beacon host; bioETH-Beacon
relaxes that assumption by protecting query and count contents from on-chain observers.

\textbf{Adjacent secure-genomics and confidential-computation constructions.}
Prior work has explored multiparty computation for secure GWAS
\citep{cho2018secure}, trusted execution environments for privacy-preserving
genomic analysis \citep{asvadishirehjini2020tee}, and decentralized private computation models
\citep{bowe2020zexe}.  These systems are not direct Beacon
count implementations, but they illustrate the broader design space for moving
genomic or application-specific computation away from plaintext centralized
processing.  The fhEVM approach is closest in spirit to TEE-based designs
because computation is delegated to a single substrate, but it replaces
hardware trust with cryptographic trust in TFHE/Ring Learning with Errors
(RLWE) and public blockchain consensus.

\textbf{Homomorphic encryption for genomics.}
Kim and Lauter \citep{kim2015private} applied HE to GWAS statistics;
Blatt \emph{et al.}\ \citep{blatt2020secure} scaled the HE pipeline to
$25{,}000$ samples; McLaren \emph{et al.}\ \citep{mclaren2016privacy}
demonstrated an HE-based HIV clinical test; Raisaro \emph{et al.}\
\citep{raisaro2019medco} built MedCo, a privacy-preserving distributed
exploration system.  iDASH competition entries \citep{wang2017idash}
catalog the design space.  These systems target richer queries than
Beacon's count primitive but assume one or more centralized compute
parties; bioETH-Beacon recovers the simplest
useful query --- the Beacon count --- without a designated evaluator.

\textbf{FHE on a programmable chain.}
The fhEVM \citep{zamafhevm,zamafhevmdocs} provides an FHE coprocessor for
Ethereum-compatible smart contracts.  Earlier blockchain-privacy and
verifiability work focused on anonymous payments \citep{sasson2014zerocash},
decentralized private computation \citep{bowe2020zexe}, or scalable transparent
computational integrity \citep{ben2018starkware}.  bioETH-Beacon explores a
complementary point in this design space: application-specific encrypted count
computation over a programmable FHE-enabled contract substrate.

\textbf{Positioning.}
Relative to MedCo \citep{raisaro2019medco}, Cho's MPC-GWAS
\citep{cho2018secure}, TEE-based genomic services
\citep{asvadishirehjini2020tee}, and McLaren et al.'s HE clinical pipeline
\citep{mclaren2016privacy}, bioETH-Beacon targets the simplest GA4GH primitive
--- the aggregate count --- and removes the designated compute evaluator
entirely: the contract performs the encrypted scan on a public blockchain
rather than at a designated compute party or inside a TEE.  The trade-off is
narrower expressiveness (aggregate count and four-way conjunction, not GWAS or
record-level queries) in exchange for a smaller compute-side trust footprint:
the only off-chain trusted component is the threshold KMS, not a compute
coordinator.  A precise latency/cost comparison against these systems is out
of scope here because their published numbers cover different query shapes;
a side-by-side benchmark on a shared workload is roadmap work.

\section{Conclusion}
\label{sec:conclusion}

We have presented bioETH-Beacon, a confidential implementation of the
GA4GH Beacon ``aggregate count'' query on a fully homomorphic EVM.
Approved hospitals upload encrypted marker-count entries; an
authorized researcher submits an encrypted marker; the contract scans
the encrypted dataset with an oblivious equality--select--add kernel and
returns an encrypted aggregate whose plaintext can be retrieved only by the
requester, via the off-chain KMS-mediated decryption path.  A
$3\!\times\!4$ tier-by-query-family grid spans the privacy/cost frontier, and a multi-filter contract collapses
the four families into a single encrypted four-way conjunctive query.
Bounded query-noise covers $4$ of the $13$ implemented contract paths
(genotype T3/T4/T5 and the G1 multi-filter); on those paths an on-chain bounded
uniform noise mechanism sampled inside the FHE coprocessor increases
anti-probing cost without reintroducing a trusted curator.  The remaining $9$
single-axis sex, age, and phenotype filter contracts currently return exact
encrypted counts to the authorized requester, so sensitive cohorts must rely on
governed requester access and rate-limit policy; formal repeated-query privacy
requires rate limiting, deduplication, or a stronger composition-aware
mechanism.

Empirical evaluation on PGS Catalog-derived panels confirms linear gas
scaling through P$100$K, distinguishes the analytic add-chain ceiling of $37$
from the largest tested passing \texttt{queryChunkSize} of $29$ and the
deployment-manifest setting of $23$, and shows that pre-aggregation cuts query
gas by $94\%$ in the fixed-$M{=}20$ tier-comparison benchmark.  Eleven security
invariants are stated across the implemented contract suite and checked with
Hardhat tests, with two Sepolia-only pending T4 ceiling probes.  The protocol provides a
practical building block for confidential genomic data sharing in
consortia where neither the host nor the requester can be assumed to be
fully trusted, and it establishes a starting point for richer Beacon-style
queries that move further into the encrypted domain.

\section*{Biographical Note}
Christos Galanopoulos, Kimon Antonios Provatas, and Ilias
Georgakopoulos-Soares are researchers at The University of Texas at Austin
studying pharmacology, genomics, and privacy-preserving biomedical
computation at the Dell Pediatric Research Institute.

\section*{Ethics Statement}
This study did not involve recruitment of human participants, intervention on
human subjects, or analysis of identifiable human participant data.  The
empirical evaluation used synthetic benchmark panels derived from public
PGS Catalog metadata and generated artifacts, so institutional review board
approval and informed consent were not required for the reported experiments.

\section*{Data Availability Statement}
No new human participant datasets were generated or analyzed in this study.
Synthetic PGS-derived benchmark panels and measurement artifacts were generated
for the empirical evaluation by the benchmark pipeline via
\texttt{npm run benchmark:panels}, \texttt{npm run profile:gas}, and
Hardhat tests.  The reported tables should be compared against
the bundled measured-run reports and generated artifacts rather than read as a
byte-for-byte guarantee for future reruns under different dependency versions.

\section*{Code Availability Statement}
The source code, smart contracts, tests, benchmark scripts, generated JSON
artifacts, reports, and figure sources supporting this manuscript are available
at \url{https://github.com/Georgakopoulos-Soares-lab/bioETH-Beacon}.
The repository represents a research prototype and should be independently
reviewed, configured, and governed before any production or clinical use.

\bibliographystyle{abbrvnat}
\bibliography{bioeth_beacon}

\end{document}